\documentclass[acmsmall,screen]{acmart}
\usepackage{booktabs}   
\usepackage{subcaption} 

\usepackage{soul}

\usepackage{array}
\newcolumntype{H}{>{\setbox0=\hbox\bgroup}c<{\egroup}@{}}

\usepackage{xr}

\usepackage{float}
\usepackage{multicol}
\usepackage{fixltx2e}
\usepackage{todonotes}
\usepackage{amsmath}
\usepackage{xspace}
\usepackage{refs}
\usepackage{galois}

\usepackage{multirow}
\usepackage[framemethod=tikz]{mdframed}

\usepackage{wrapfig}
\usepackage[export]{adjustbox}
\usepackage{stfloats}

\newcommand{\OnlyConference}[1]{}
\newcommand{\OnlyTech}[1]{}
\newcommand{\OnlySupplementary}[1]{}
\newcommand{\OnlySupplementaryPrologue}[1]{}

\newcommand{\FuzzOut}[1]{}

\usepackage{tikz}
\usetikzlibrary{shapes}
\usetikzlibrary{patterns}
\usetikzlibrary{circuits}
\usetikzlibrary{fit}
\usetikzlibrary{matrix}
\usetikzlibrary{arrows}
\usetikzlibrary{calc}
\usetikzlibrary{positioning}
\tikzset{>=stealth',every on chain/.append style={join}, every join/.style={->}}

\usepackage{makecell}

\usepackage{etoolbox}

\newrobustcmd*{\nsquare}[1]{\tikz{\filldraw[draw=#1,fill=#1] (0,0)
    rectangle (0.2cm,0.2cm);}}

\newrobustcmd*{\ncircle}[1]{\tikz{\filldraw[draw=#1,fill=#1] (0,0) circle [radius=0.1cm];}}

\newrobustcmd*{\ntriangle}[1]{\tikz{\filldraw[draw=#1,fill=#1] (0,0) --
    (0.2cm,0) -- (0.1cm,0.2cm);}}

\usepackage{listings}
\usepackage{listingsutf8}

\definecolor{darklava}{rgb}{0.58, 0.1, 0.9}

\usepackage{comment}
\usepackage[algo2e,linesnumbered,ruled,vlined,resetcount]{algorithm2e}
\usepackage{textcomp}

\SetKwComment{Comment}{$\triangleright$\ }{}

\SetCommentSty{xCommentSty}
\usepackage{color, colortbl}
\usepackage{xifthen}

\newcommand{\tool}{\textsc{Amurth}\xspace}
\newcommand{\func}[1]{\textsc{#1}}
\newcommand{\abst}[2]{#1^\sharp_{#2}}
\newcommand{\pex}{E^{+}}
\newcommand{\nex}{E^{-}}

\newcommand{\fsyn}[1][]{$f_{E}^\sharp$}
\newcommand{\fSharpSyn}[1][]{
  \ifthenelse{\isempty{#1}}
             {f_{E}^\sharp}
             {f_{E}^\sharp\left{#1}\right}
}

\newcommand{\subsubsubsection}[1]{\vspace{2pt plus 1pt minus 1pt}\noindent{\bf #1}}
\newcommand{\AbsDomain}{A}

\newcommand{\PowerSet}[1]{\mathcal{P}(#1)}

\newcommand{\safestr}{SAFE\textsubscript{str}\xspace}
\newcommand{\csdom}{\mathcal{CS}}
\newcommand{\ssk}{\mathcal{SS}_k}
\newcommand{\charin}{\mathcal{CI}}
\newcommand{\psdom}{\mathcal{PS}}
\newcommand{\intv}{\mathcal{A}_{intv}}
\newcommand{\sintv}{\mathcal{A}_{\mathtt{sintv}}}
\newcommand{\uintv}{\mathcal{A}_{\mathtt{uintv}}}
\newcommand{\wintv}{\mathcal{W}}

\newcommand{\sq}{\textquotesingle}

\newcommand{\bestset}{\mathcal{\widehat{S}}_L}

\usepackage{enumitem}
\setlist{leftmargin=*}
\setlist{nosep,leftmargin=\parindent}
\setlist[description]{leftmargin=\parindent,labelindent=\parindent}

\definecolor{crimson}{HTML}{C90016}
\definecolor{tmagenta}{HTML}{FF00FF}
\definecolor{tred}{HTML}{FF0000}
\definecolor{posex}{HTML}{0ca765}
\definecolor{bottlegreen}{rgb}{0.0,0.42,0.31}

\newcommand{\loris}[1]{{\color{magenta}{L: #1}}}
\newcommand{\twr}[1]{{{\color{purple}{T: #1}}}}
\newcommand{\subhajit}[1]{{{\color{bottlegreen}{S: #1}}}}
\newcommand{\sujit}[1]{{\color{green!70!blue}{SJT: #1}}}
\newcommand{\pankaj}[1]{{\color{crimson}{PK: #1}}}

 \renewcommand{\loris}[1]{}
 \renewcommand{\twr}[1]{}
 \renewcommand{\subhajit}[1]{}
 \renewcommand{\pankaj}[1]{}
 \renewcommand{\sujit}[1]{}

\newcommand{\forARXIV}[1]{#1}
\newcommand{\forOOPSLA}[1]{#1}

\renewcommand{\forOOPSLA}[1]{}

\newcommand{\twrchanged}[1]{{{\color{cyan}{#1}}}}

\newcommand{\skmchanged}[1]{\textcolor{green!45!black}{#1}}

\renewcommand{\twrchanged}[1]{#1}

\renewcommand{\skmchanged}[1]{1}

\newcommand{\Omit}[1]{}
\newcommand{\B}[1]{\langle{#1}\rangle}
\newcommand{\pre}{\textit{pre}}
\newcommand{\suf}{\textit{suf}}
\newcommand{\lcp}{\textit{lcp}}
\newcommand{\lcs}{\textit{lcs}}
\newcommand{\best}{{\widehat{f}^{\,\sharp}}}
\newcommand{\bestl}{{\widehat{f}^{\,\sharp}_L}}
\newcommand{\beste}{{\widehat{f}^{\,\sharp}_E}}
\newcommand{\abstf}{\abst{f}{}}
\newcommand{\abstfe}{\abst{f}{E}}

\newcommand{\csclosed}{\textsc{CheckSoundness}^{\mathcal{R}un}}
\newcommand{\cslogic}{\textsc{CheckSoundness}^{\Phi}}

\newcommand{\eqdef}{\operatorname{\,{=_{\textit{df}}}\,}}

\AtBeginDocument{%
  \providecommand\BibTeX{{%
    \normalfont B\kern-0.5em{\scshape i\kern-0.25em b}\kern-0.8em\TeX}}}

    \acmConference[PL'18]{ACM SIGPLAN Conference on Programming Languages}{January 01--03, 2018}{New York, NY, USA}
    \acmYear{2018}
    \acmISBN{} 
    \acmDOI{} 
    \startPage{1}

    \setcopyright{none}

\begin{document}

\title{Synthesizing Abstract Transformers}


\author[Kalita]{Pankaj Kumar Kalita}
\affiliation{%
  \institution{Indian Institute of Technology Kanpur}
  \state{Uttar Pradesh}
  \country{India}}
\email{pkalita@cse.iitk.ac.in}

\author[Muduli]{Sujit Kumar Muduli}
\affiliation{%
  \institution{Indian Institute of Technology Kanpur}
  \state{Uttar Pradesh}
  \country{India}
}
\email{smuduli@cse.iitk.ac.in}

\author[D'Antoni]{Loris D'Antoni}
\affiliation{%
 \institution{University of Wisconsin--Madison}
 \country{USA}}
\email{ldantoni@wisc.edu}

\author[Reps]{Thomas Reps}
\affiliation{%
  \institution{University of Wisconsin--Madison}
  \country{USA}}
\email{reps@cs.wisc.edu}

\author[Roy]{Subhajit Roy}
\affiliation{%
  \institution{Indian Institute of Technology Kanpur}
  \state{Uttar Pradesh}
  \country{India}
}
\email{subhajit@cse.iitk.ac.in}





\begin{abstract}
This paper addresses the problem of creating abstract transformers
automatically.
The method we present automates the construction of static analyzers
in a fashion similar to the way \texttt{yacc} automates the
construction of parsers.
Our method treats the problem as a program-synthesis problem.
The user provides specifications of
(i) the concrete semantics of a given operation $\textit{op}$,
(ii) the abstract domain $\AbsDomain$ to be used by the analyzer, and
(iii) the semantics of a domain-specific language $L$ in which the abstract
transformer is to be expressed.
As output, our method creates an abstract transformer for $\textit{op}$ in
abstract domain $\AbsDomain$, expressed in $L$ (an ``$L$-transformer for
$\textit{op}$ over $\AbsDomain$'').
Moreover, the abstract transformer obtained is a \emph{most-precise}
$L$-transformer for $\textit{op}$ over $\AbsDomain$;
that is, there is no other $L$-transformer for $\textit{op}$ over $\AbsDomain$
that is strictly more precise.

We implemented our method in a tool called \tool.
We used \tool to create sets of replacement abstract transformers for
those used in two existing analyzers, and obtained essentially
identical performance.
However, when we compared the existing transformers with the
transformers obtained using \tool,
we discovered that four of the existing transformers were unsound,
which demonstrates the risk of using manually created transformers.
\end{abstract}




\renewcommand\footnotetextcopyrightpermission[1]{} 
\fancyfoot{}
\makeatletter
\let\@authorsaddresses\@empty
\makeatother

\maketitle

\section{Introduction}
\label{Se:Introduction}

Abstract interpretation is a methodology for establishing whether
a software system satisfies desired properties.
It obtains information about the states that a program
(possibly) reaches during execution, without actually running
the program on specific inputs.
Instead, the program's behavior is explored for all possible inputs,
and all possible states that the program can reach, by running the program
over \emph{abstract values}---descriptors that represent sets of states.
Each operation of the program is interpreted over abstract values
in a manner that overapproximates the operation's standard
(``concrete'') interpretation over the corresponding
sets of concrete states.
Such an interpretation of operation $\textit{op}$ is called
an \emph{abstract transformer} for $\textit{op}$ (denoted by
$\textit{op}^\sharp$).
In particular, the result of applying an abstract transformer for a
statement must result in an abstract value that represents a superset
of the concrete states that can actually arise.

\citet{POPL:CC77} showed that, under reasonable conditions,
for each operation $\textit{op}$, there exists a \emph{most-precise
abstract transformer} $\textit{op}^\sharp$ (or ``best transformer for $\textit{op}$'').
When the power set of concrete states $\PowerSet{C}$
is related to the set of abstract values (\emph{abstract domain} $\AbsDomain$)
by a Galois connection $\PowerSet{C} \galois{\alpha}{\gamma} \AbsDomain$,
the best transformer for $\textit{op}$ is the function
\begin{equation}
  \label{Eq:BestTransformerSpecification}
  \abst{\widehat{\textit{op}}}{} = \alpha \circ \widetilde{\textit{op}} \circ \gamma,
\end{equation}
where $\widetilde{\textit{op}}$ is the lifting of $\textit{op}$ to sets of
concrete states.
Informally,
$\widehat{\textit{op}}$ runs $\textit{op}$ on all the
states represented by the input abstract value.
\eqref{BestTransformerSpecification} defines the limit of
precision obtainable using
abstraction function $\alpha$ and concretization function $\gamma$.
However, \eqref{BestTransformerSpecification} is just a
\emph{specification} of the best transformer:
it fails to provide an \emph{algorithm} for either
\begin{enumerate}[label=\color{blue}(\alph*),leftmargin=1.5\parindent]
  \item
    \label{It:ApplyingTheBestTransformer}
    applying $\abst{\widehat{\textit{op}}}{}$ to a given abstract value, or
  \item
    \label{It:RepresentingTheBestTransformer}
    finding a representation of $\abst{\widehat{\textit{op}}}{}$.
\end{enumerate}
A ``representation'' means either (i)~a data structure
whose interpretation is $\abst{\widehat{\textit{op}}}{}$, or 
(ii)~a program to perform $\abst{\widehat{\textit{op}}}{}$.

Prior work on algorithms for
\ref{It:ApplyingTheBestTransformer} and
\ref{It:RepresentingTheBestTransformer} can be categorized as follows.
For \ref{It:ApplyingTheBestTransformer}, \citet{CAV:GS97}
showed that for \emph{predicate-abstraction domains} (i.e,
domains based on a fixed, finite set of state predicates), SMT solvers
can be used to apply the best transformer to an abstract value.
(Improved techniques were given by \citet{CAV:LBC05,CAV:LNO06}.)
\citet{VMCAI:RSY04} showed that for \emph{abstract
domains with no infinite ascending chains}, SMT solvers could be
used to apply the best transformer to an abstract value.
The drawback of these methods is that they generally require
making a large number of SMT calls---at least one SMT call for
each used abstract value.

For \ref{It:RepresentingTheBestTransformer},
\citet{PLDI:SLC07} gave a method for creating a
representation of an abstract transformer for predicate-abstraction domains.
Because their method was based on term-rewriting heuristics, they had
no guarantee of obtaining a representation of the best abstract transformer.
\citet{ELSAR14} gave a method for creating a representation
of a best abstract transformer for the abstract domain of
\emph{conjunctions of bit-vector equalities}.
That method uses calls on an SMT solver at the time a transformer
is created for a statement, basic-block, or large-block encoding \citep{FMCAD:BCGKS09};
thereafter, all operations are carried out within the abstract domain.

The advantage of methods of type
\ref{It:RepresentingTheBestTransformer} is that they compile the
abstract transformer to a form that can be used---i.e., applied or
composed---without further incurring any expensive operations, such as
SMT calls;
all expensive operations are performed once and for all at \textit{compilation time}.

Thanks to our synthesis-based approach---and in contrast with previous
work---the work described in this paper is not limited to
predicate abstraction or bit-vector equalities.
However, our work addresses a slightly different problem from prior work:
our method is parameterized by a domain-specific language (DSL) $L$
in which the abstract transformer for operation $\textit{op}$ is to
be expressed.
Given $\textit{op}$ and abstract domain $\AbsDomain$,
our method creates an abstract transformer for $\textit{op}$ over
$\AbsDomain$, \emph{expressed in DSL $L$}---what we call
an ``$L$-transformer (for $\textit{op}$ over $\AbsDomain$).''
Our algorithm is guaranteed to return a \emph{best} $L$-transformer.
That is, among all $L$-transformers for $\textit{op}$ over $\AbsDomain$,
there is no other $L$-transformer that is strictly more precise than the
one obtained by our algorithm.
(There may be other $L$-transformers that are incomparable to the one
obtained by our algorithm, which is why we say that the algorithm creates
\emph{a} best $L$-transformer.)

\FuzzOut{
Our framework supports two usage scenarios:
      \begin{itemize}
        \item
          In the first, a logical specification of
          the behavior of concrete operator $\textit{op}$ is available.
          In this case, our technique is able to ``compile'' the logical
          specification into abstract transformers for multiple abstract domains
          for use with abstract-interpretation engines.
								\FuzzOut{In the second, only an operational specification of $\textit{op}$ is available.
		  In this case, which is a common real-world verification scenario where certain
          operations are only available as external library routines (e.g., for which one only has the implementation in machine code), our technique
		  is still able to generate transformers. 
		  Moreover, this capability can relieve verification engineers from the daunting task of having to 
											provide ``stubs,'' over-approximating summaries, or mock implementations for library code.}
      \end{itemize}
Thanks to our automated synthesis technique, we discovered four soundness bugs
in the manually written transformers of two real-world abstract-interpretation
frameworks\FuzzOut{---even with only closed-box access to the concrete operations}.
}

Thanks to our automated synthesis technique, we discovered four soundness bugs
in the manually written transformers of two real-world
abstract-interpretation frameworks.

\subsubsubsection{Contributions.}
We present a framework to create abstract transformers in a form
in which the application of an abstract transformer involves the
execution of relatively simple code.
In particular, we advance the type
\ref{It:RepresentingTheBestTransformer} approach in two ways:
\begin{itemize}
  \item
    Our framework treats the problem as a
    \emph{program-synthesis problem}, which opens up a new possibility:
    rather than the representation of $\abst{\widehat{\textit{op}}}{}$
    being limited to a fixed interface of operations provided by
    an abstract domain,
    the user can supply a DSL---by specifying both its syntax and semantics---in which
    $\abst{\widehat{\textit{op}}}{}$ is to be expressed.
 
  \item
    The assumptions of the framework are fairly minimal.
    The inputs are quite natural (\sectref{Overview}).
    The synthesis algorithm that serves as the engine of the framework
    (\sectrefs{soundvsprecise}{Algorithm}) is formalized using
    a primitive \func{Synthesize} for synthesizing candidate $L$-transformers;
    a primitive \func{MaxSatSynthesize}, which is biased toward satisfying
    so-called ``positive'' examples---see \sectref{soundvsprecise}; and
    two primitives that check soundness and precision of candidate $L$-transformers,
    generating counterexamples when the respective property fails to hold.

\FuzzOut{
  \item 
				Based on the usage scenarios given above, the oracle to check soundness has two different modes of operation: \begin{itemize}
        \item
          When the logical specification is available, soundness is checked via automated-reasoning methods.
        \item
					When only an operational specification of $\textit{op}$ is available, soundness is checked in an approximate manner, using fuzzing.
      \end{itemize}
      Our experiments show that our method works well in both modes,
      including the second, approximate, mode.
}

  \item
    We implemented a tool, called \tool (\sectref{Implementation}),
    to support our framework, and obtained good results:
    we used \tool to create sets of replacement abstract
    transformers for those used in two existing analyzers,
    and obtained essentially identical performance (\sectref{Experiments}).
    However, when we compared the existing transformers with the
    replacements synthesized by \tool,
    we discovered that four of the existing transformers were \textit{unsound}.
    These results demonstrate the risk of using manually created transformers,
    and hence the value of a tool for creating them automatically.
\end{itemize}


\forOOPSLA{
\fbox{\parbox{.93\textwidth}{
  \textbf{For additional details, see the extended version of this work~\cite{amurthARXIV}.}
}}
}


\section{Problem Statement}
\label{Se:Overview}

In this section, we define the problem addressed by our framework.
Throughout the paper, we use a running example in which the goal is to
synthesize a most-precise $L$-transformer for the absolute-value
function $\mathtt{abs(x) = |x|}$ over the domain of intervals, where
$L$ is the DSL defined by
\begin{equation}
  \label{Eq:LForIntervalDomain}
  \begin{array}{@{\hspace{0ex}}r@{\hspace{1.0ex}}c@{\hspace{1.0ex}}l@{\hspace{0ex}}}
    \textit{Transformer} & ::= & \lambda \texttt{a} . [E, E] \\
                       E & ::= & \texttt{a.l} \mid \texttt{a.r} \mid  0 \mid {-}E \mid {+}\infty \mid {-}\infty \mid E + E \mid E - E \mid E * E \mid \texttt{min}(E, E) \mid \texttt{max}(E, E)
  \end{array}
\end{equation}
and the operations in $E$ have their standard meaning.
We describe what a user of the framework has to provide to solve this problem,
and what they obtain as output.




The user of our framework needs to provide the following inputs:

\noindent
{\bf Concrete domain:}
    A definition of the concrete domain $C$---typically some set of values or program states.
    In our example, $C$ is the set of integers.

\noindent
{\bf Concrete transformer:}
    A definition of the concrete semantics of the function $f$ for
    which we are trying to synthesize a most-precise $L$-transformer.
    In our example, $f$ is the function $\texttt{abs}: \texttt{Int} \rightarrow \texttt{Int}$,
    which takes as input an integer and returns its absolute value.
    The semantics of $\texttt{abs}(x)$ is provided by a logical specification
    $\Phi_{\mathtt{abs}}(x,x') \eqdef (x \ge 0 \land x' = x) \lor (x < 0 \land x' = -x)$,
    where $x$ and $x'$ represent the input and output, respectively.

\noindent
{\bf Abstract domain:}
    A definition of the abstract lattice $(A, \sqsubseteq, \bot)$, where $A$ is the abstract domain,
    $\sqsubseteq$ is the partial order on elements of $A$, and
    $\bot$ is the least element of $A$.
    In our example, the domain of \emph{intervals} $\intv$ abstracts a set of integers by maintaining only the maximum and minimum elements in the set.
    Each element $a$ is a pair $[a.l,a.r]$ such that $a.l$ (which can be $-\infty$) denotes the minimum element
    and
    $a.r$ (which can be $+\infty$) denotes the maximum element.

\noindent
{\bf Relation between the abstract and concrete domains:}
    A definition of the concretization function $\gamma : A \rightarrow \PowerSet{C}$.
    In our example, the concretization function is $\gamma(a) = \{a.l, a.l+1, \dots, a.r\}$.

\noindent
{\bf Language of Possible Transformers:}
    The syntax and semantics of a DSL in which the synthesizer is to express
    abstract transformers.
    In our example, the DSL defined in \eqref{LForIntervalDomain}.

We assume that all semantic specifications are given in---or can be translated to---formulas
in a fragment of first-order logic.
For instance, the concretization function for intervals, $\gamma_{\textit{Interval}}(\mathtt{a})$,
can be specified via the predicate
$x \in \gamma_{\textit{Interval}}(\mathtt{a}) \eqdef \mathtt{a.l} \le x \land x \le \mathtt{a.r}$,

For concrete domain $C$ and abstract domain $A$, \eqref{BestTransformerSpecification} specifies the
behavior of the best abstract transformer for $f$, denoted by $\abst{\widehat{f}}{}$.
As mentioned previously, \eqref{BestTransformerSpecification} does not
provide the basis for an implementation of $\abst{\widehat{f}}{}$
because $\gamma(a)$ is potentially a large set.
(It can even be an infinite set for some abstract domains.)
Moreover, the introduction of language $L$ into the problem introduces
a new wrinkle: there is no guarantee that $\abst{\widehat{f}}{}$ is
even expressible in language $L$.


%

Any transformer that is expressible in $L$ is referred to as an \textit{L-transformer},
denoted by $\abst{f}{L}$.
We use $\abst{\widehat{f}}{L}$ to denote a \emph{best} $L$-transformer for $f$. 
A best $L$-transformer must satisfy the dual objectives of soundness and precision.

\noindent
{\bf Soundness:} A \textit{sound} $L$-transformer for a concrete function $f$ must overapproximate the best transformer $\abst{\widehat{f}}{}$;
        i.e., $\abst{f}{L}$ is sound iff for all $a \in A$,
        $ \best(a) \sqsubseteq \abst{f}{L}(a).$

\noindent
{\bf Precision:} We define a (pre-)partial order on $L$-transformers with respect to precision ($\sqsubseteq_{pr}$) as follows: for all $\abst{{f_1}}{}, \abst{{f_2}}{} \in L$,
      $ \abst{{f_1}}{} {\sqsubseteq}_{pr} \abst{{f_2}}{} \equiv \forall a\in A.\ \gamma(\abst{f}{1}(a)) \subseteq \gamma(\abst{f}{2}(a))$. A sound $L$-transformer $\abstf \in L$ is \textit{most-precise} if it is minimal with respect to ${\sqsubseteq}_{pr}$.

\begin{definition}
  An abstract transformer $\abst{f}{L} \in L$ is \textbf{\emph{a best
  $L$-transformer}} for a function $f$ if $\abst{f}{L}$ is both sound
  and most-precise (in which case, we denote it by $\bestl$).
  We use $\bestset(f)$ to denote the set of all best $L$-transformers for a function $f$.
\end{definition}

Note that there may not exist a unique best $L$-transformer under $\sqsubseteq_{pr}$.
For example, if $f$ is the constant-zero function $\lambda x . 0$, and language $L$
can only express the functions $\{\lambda a.[0,k], \lambda a.[-k, 0] \mid k\in \mathbb{N} \wedge k\geq 1 \}$,
the transformers $\lambda a.[0,1]$ and $\lambda a.[-1,0]$, which are
incomparable under $\sqsubseteq_{pr}$, are both best $L$-transformers
for $f$.

%

This paper targets the following problem:
\begin{mdframed}[innerleftmargin = 3pt, innerrightmargin = 3pt, skipbelow=-0.25em]
  Given the concrete semantics $\Phi_f$ of a concrete transformer $f$, a description of an abstract domain ($A, \sqsubseteq, \sqcup$), its relation to the concrete domain ($\gamma$), and a domain-specific language $L$, synthesize a best $L$-transformer for $f$. 
  \end{mdframed}

\noindent

As illustrated in \sectref{Algorithm}, our method synthesizes the following
transformer $\mathtt{\abst{abs}{} : \intv \rightarrow \intv}$:
\begin{equation}
  \label{Eq:BestTransformerExpressionForAbs}
  \mathtt{\abst{abs}{}(a) = [max(max(0, a.l), -a.r), \ max(-a.l, a.r)]}.
\end{equation}
Even though the concrete function \texttt{abs} and the interval
abstract domain are both quite simple, the transformer in
\eqref{BestTransformerExpressionForAbs} is non-trivial, providing
motivation for this work.


We now provide an informal argument that the transformer $\mathtt{abs}^\sharp$ is
a best $L$-transformer $\abst{\widehat{\textit{abs}}}{L}$ over the interval domain.
Given an input interval $\texttt{a} \in \mathcal{A}_{\textit{intv}}$,
$\abst{\widehat{\textit{abs}}}{L}$
must behave as follows:
(1) If $\gamma(\texttt{a})$ only contains non-negative values (i.e.,
$\mathtt{a.l} \geq 0$),
$\abst{\widehat{\textit{abs}}}{L}$
should return the input interval \texttt{a} itself.
(2) If $\gamma(\texttt{a})$ only contains non-positive values (i.e.,
$\mathtt{a.r} \leq 0$),
$\abst{\widehat{\textit{abs}}}{L}$
should return the interval $\mathtt{[-a.r, -a.l]}$.
(3) If $\gamma(\texttt{a})$ contains both positive and negative values
(i.e., $\texttt{a.l} < 0 \land \texttt{a.r} > 0$),
$\abst{\widehat{\textit{abs}}}{L}$
should return the interval $\mathtt{[0, max(-a.l, a.r)]}$.
A transformer meeting all these conditions is a best $L$-transformer (cf.\ \eqref{BestTransformerSpecification}).
With a little bit of case analysis, one can see that the transformer $\mathtt{abs}^\sharp$
handles all of the cases as described above.

Note that, for a given abstract domain, all of the components provided
as inputs to the framework are reusable.
To synthesize a best $L$-transformer for a different concrete
transformer $g$, one only needs to supply the specification of $g$.


For a given concrete transformer $f$, to synthesize a variety of best $L_i$-transformers
over different abstract domains $(A_i, \sqsubseteq_i, \bot_i)$, one needs
to supply the definitions of the different abstract domains,
and---typically---define DSLs $L_i$ with suitable operations to manipulate
the various components in the representations of $A_i$ values.


\section{Positive Examples, Negative Examples, Soundness, and Precision}
\label{Se:soundvsprecise}

The engine that underlies our framework is
an example-based synthesis algorithm to synthesize the target $L$-transformer.
\twrchanged{
The key insight behind the algorithm is as follows:
}

\twrchanged{
\fbox{\parbox{.93\textwidth}{
  Use both positive and negative examples.
  Treat positive examples as hard constraints and negative examples as soft constraints.
}}
}

\noindent
\twrchanged{
The synthesis algorithm is formalized using a primitive \func{Synthesize} for synthesizing
candidate $L$-transformers;
a primitive \func{MaxSatSynthesize}, which is biased toward satisfying
positive examples; and
two primitives, \func{CheckSoundness} and \func{CheckSoundness}, which
check the soundness and precision of candidate $L$-transformers,
respectively---generating counterexamples when the respective property
fails to hold.
}
The algorithm proceeds in iterations, and maintains a set of examples $E = \B{\pex,\nex}$,
divided into positive ($\pex$) and negative ($\nex$) examples.
It also issues \emph{queries} to check whether the current candidate $L$-transformer
is sound and precise.
If it fails either the soundness or precision criterion, a new candidate $L$-transformer
is created.
\twrchanged{
In this section, we discuss the soundness and precision queries (\sectrefs{CheckSoundness}{CheckPrecisionStar}, respectively).
The algorithm itself is presented in~\sectref{Algorithm}, and the role of
\func{MaxSatSynthesize} is explained in \sectref{ConsistencyOfPositiveAndNegativeExamples}.
}

\begin{definition}[Positive and Negative Examples]
    A \emph{positive example} is a pair $\B{a,c'}$ such that $a \in A$ and $c' \in \gamma(\best(a))$.
    A \emph{negative example} is a pair $\B{a,c'}$ such that $a\in A$, and there exists \emph{some} best $L$-transformer $\bestl\in \bestset$ such that $c' \notin \gamma(\bestl(a))$.
\end{definition}

\begin{example}
For the interval domain and the function $\texttt{abs}$,
$\langle [5, 9], 6\rangle$ is a positive example, but $\langle [5, 9], 12\rangle$ is not. 
Along the same lines, assuming the DSL $L$  from \eqref{LForIntervalDomain},
$\langle [5,12], 2\rangle$ is a negative example, while $\langle [5,12], 7\rangle$ is not a negative example.
\end{example}

\begin{wrapfigure}{R}{.3\linewidth}
  \centering
  \includegraphics[frame,scale=0.4]{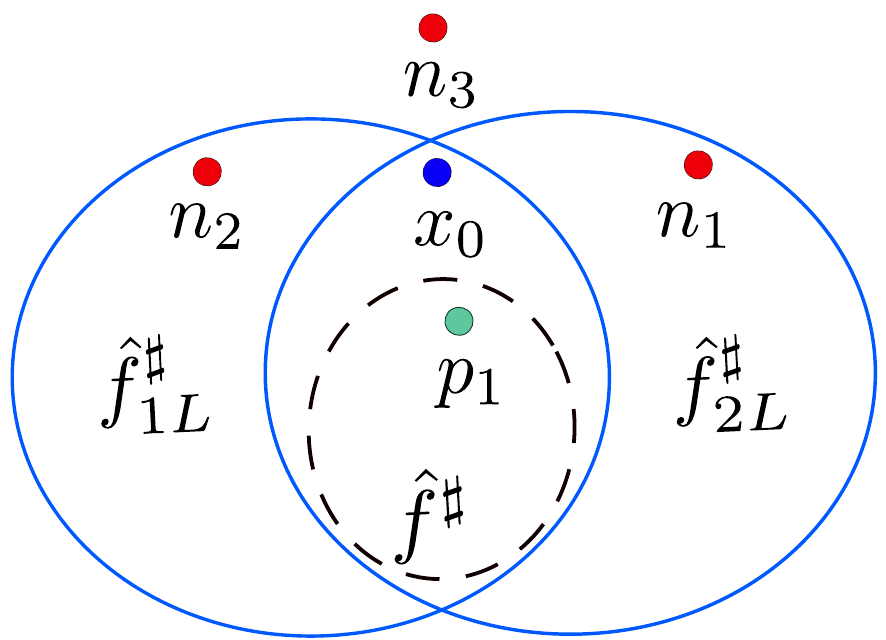}
  \caption{\label{Fi:bunny}
    {\color{ForestGreen}Positive} and {\color{red}negative} examples, along with two best $L$-transformers and $\abst{\hat{f}}{}$.
    The \textcolor{blue}{blue} example $x_0$ is neither positive nor negative.
  }
  \vspace{-4.0ex}
\end{wrapfigure}

\figref{bunny} illustrates a case with two best $L$-transformers,
$\bestset = \{\abst{\widehat{f}}{1L}, \abst{\widehat{f}}{2L}\}$,
shown by the {\color{blue}blue} outlines; the black dashed outline shows the best abstract transformer $\abst{\widehat{f}}{}$.
Points on the plot depict examples $\langle a, c'\rangle$;
a point $\B{a,c'}\in \abstf$ denotes that $c'\in \gamma(\abstf(a))$ and vice-versa.
Point $p_1$  is a positive example because it is inside $\abst{\widehat{f}}{}$.
Point $n_1$ (resp. $n_2$) is a negative example because it is outside best $L$-transformer $\abst{\widehat{f}}{1L}$ (resp. $\abst{\widehat{f}}{2L}$); and $n_3$ is a negative example because it is outside of both $\abst{\widehat{f}}{1L}$ and $\abst{\widehat{f}}{2L}$. Point $x_0$ is neither a positive example nor a negative example:
it is outside $\abst{\widehat{f}}{}$, but inside both best $L$-transformers.


We assume that we have available a \func{Synthesize} procedure
that accepts a set of examples $\B{\pex, \nex}$, and returns \textit{an}
$L$-transformer $\abstfe$ that includes all $e^+\in \pex$ and excludes
all $e^-\in\nex$.
That is, $\abst{f}{E}$ satisfies
\begin{equation}
  \label{Eq:satsynth}
  \begin{array}{l@{\hspace{0.5ex}}r@{\hspace{0.5ex}}l}
    \textit{sat}^+(\abst{f}{E}, \pex) \land \textit{sat}^-(\abst{f}{E}, \nex), 
         & \textrm{where}~& \textit{sat}^+(\abst{f}{E}, \pex) =_{\textit{df}} \forall \B{a,c} \in \pex \, .~c \in \gamma(\abst{f}{E}(a)),  \\
         & \textrm{and}   & \textit{sat}^-(\abst{f}{E}, \nex) =_{\textit{df}} \forall \B{a,c} \in \nex \, .~c \not\in \gamma(\abst{f}{E}(a)).
  \end{array}
\end{equation}


\def\posEx{$\langle a, c\rangle$~}
\subsection{Soundness Queries (\func{CheckSoundness})}
\label{Se:CheckSoundness}

\begin{definition}
   \label{De:SoundnessOracle}
    A \textit{soundness query} takes as input an $L$-transformer $\abstfe$ (that is consistent with the set of examples $E$) and returns
\begin{enumerate}
    \item $\mathit{True}$ if $\abstfe$ is sound for all possible inputs;
          i.e., $\forall a.\ \best(a) \sqsubseteq \abstfe(a)$,
    \item $\mathit{False}$ and a pair of abstract and concrete values $\B{a,c'}$, such that $c'\in \gamma(\best(a)) \setminus \gamma(\abstfe(a))$.
%
$a$ is a witness to the unsoundness of $\abstfe$;
$\langle a, c'\rangle$ is called a \textbf{\emph{positive counterexample}}.
\end{enumerate}
\end{definition}

Because a logical specification $\Phi_f$ for the semantics of the
concrete function $f$ is provided, where $\Phi_f$ is expressed in a
decidable logic, the soundness check can be carried out by checking
the following formula for satisfiability:
\begin{equation}
    \label{Eq:checkSoundConditionLogic}
    \exists \B{a,c'},~ \textrm{where } a \in A, \textrm{ and } c' \in C, \textrm{ such that }
    \exists c \in C, c \in \gamma(a) \land \Phi_f(c,c') \land c' \notin \gamma(\abst{f}{E}(a)) 
\end{equation}
Let us now define the interface:
\begin{equation}
    \label{Eq:checksoundness}
    {\func{CheckSoundness}}(\abstfe, f) 
        = \begin{cases}
            \textit{False}, \B{a,c'}&\mbox{if \eqref{checkSoundConditionLogic} is SAT}\\
            \textit{True}, \_&\mbox{otherwise}
          \end{cases}
\end{equation}

One might wonder if is is possible to solve the  problem of synthesizing a best $L$-transformer
using $\func{CheckSoundness}$ alone.
For example, one could use a counterexample-guided inductive synthesis (CEGIS) algorithm that
uses $\func{CheckSoundness}$ iteratively, to synthesize a succession of candidate
$L$-transformers that cover larger and larger sets of examples.
This (hypothetical) algorithm would maintain a set of positive examples $\pex$,
use \func{Synthesize} to generate a sound $L$-transformer $f^\sharp_{\pex}$ for $\pex$, and
issue a query to $\func{CheckSoundness}$ to determine whether the current
candidate $L$-transformer is sound in general.
If not, the algorithm would add the positive counterexample to $\pex$ and repeat.

\begin{figure}
	\begin{minipage}[t]{.5\linewidth}
		\centering
		\includegraphics[height=0.5\linewidth]{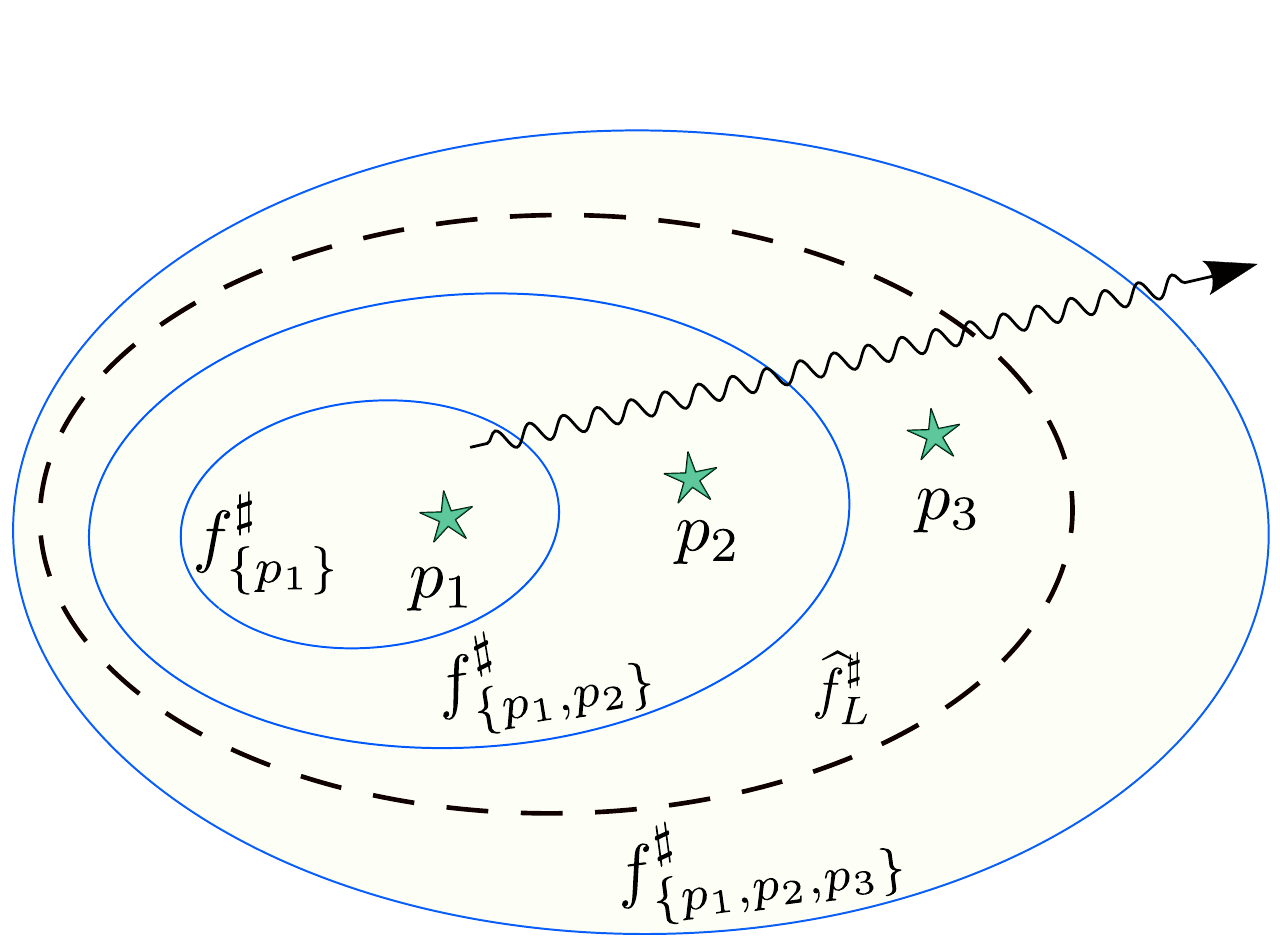}
		\subcaption{Adding positive counterexamples}
		\label{Fi:expanding-transformer}
	\end{minipage}
\hfill
	\begin{minipage}[t]{.5\linewidth}
		\centering
		\includegraphics[height=0.5\linewidth]{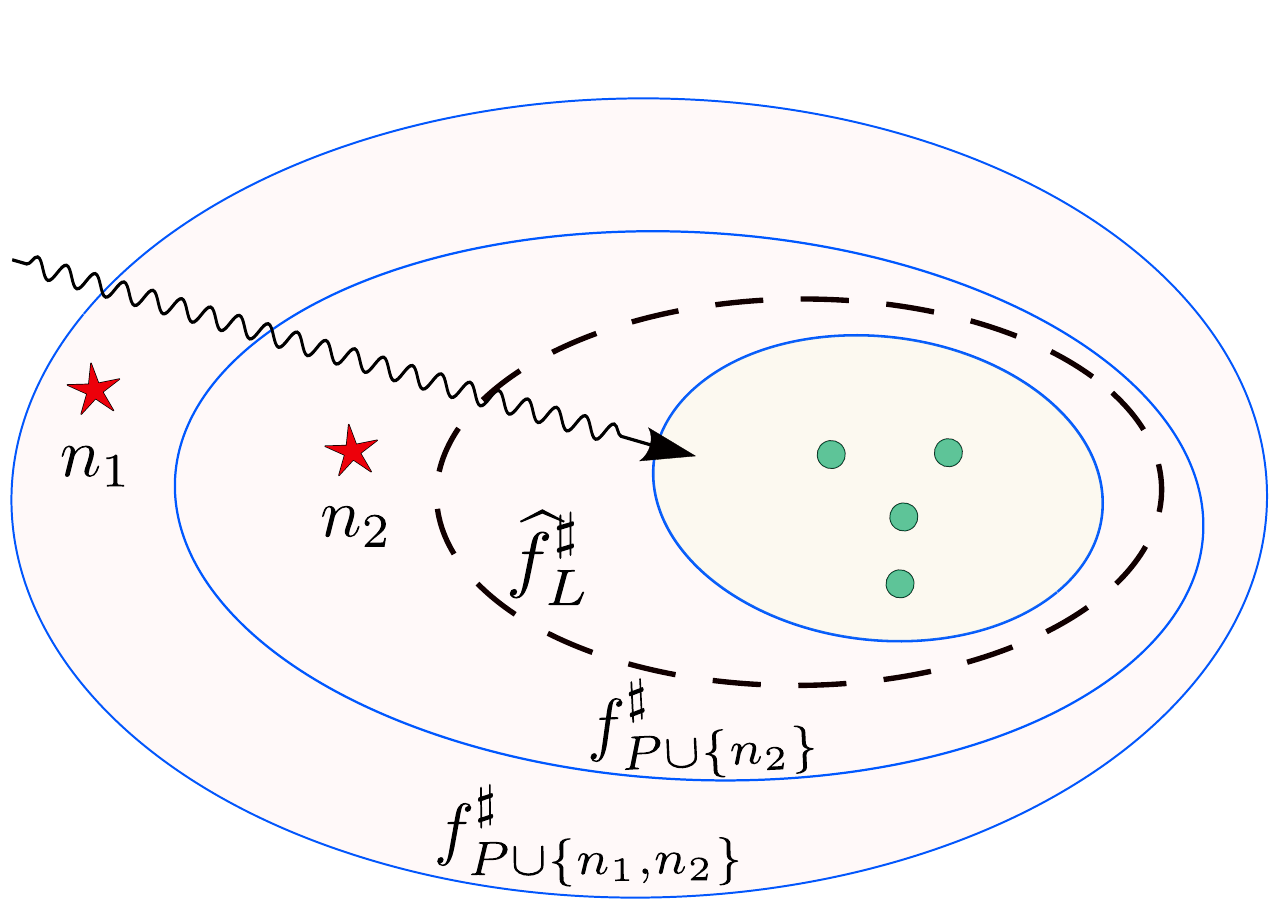}
		\subcaption{Adding negative counterexamples. $P$ is a set of positive examples (\textcolor{green!70!blue}{$\bullet$}) .}
		\label{Fi:shrinking-transformer}
	\end{minipage}
	\caption{\label{Fi:shrink-expand}The blue ovals represent the successions of synthesized $L$-transformers; the black dashed oval represents a best $L$-transformer.}
\end{figure}

For example, suppose that on some iteration
$ \abst{\texttt{abs}}{\pex_0}  = \lambda \texttt{a : [a.l, a.r]}$,
and a call to $\func{CheckSoundness}$ generates the positive counterexample
$\langle [10,15], 12 \rangle$.
A synthesizer given $\pex_1 = \pex_0 \cup \{ \langle [10,15], 12 \rangle \}$
may generate the $L$-transformer $\abst{\texttt{abs}}{\pex_1}  \mathtt{(a) = [0, a.r]}$,
which is sound on all examples.
The positive-counterexample-guided-synthesis algorithm is illustrated in \figref{expanding-transformer}.
The blue ovals represent the succession of synthesized $L$-transformers;
the black dashed oval represents the best transformer.
The green stars represent positive counterexamples.
If this algorithm terminates, it is guaranteed to generate an
$L$-transformer that is sound on all of the possible inputs;
however, the result is not guaranteed to be precise.
For instance, our example may converge to the maximally imprecise result
$\abst{\texttt{abs}}{\pex}\mathtt{(a) = [-\infty, +\infty]}$.

\subsection{Precision Queries (\func{CheckPrecision})}
\label{Se:CheckPrecisionStar}

To help the reader's understanding, we start by discussing a slightly idealized version
of the \func{CheckPrecision} query, denoted by \func{CheckPrecision$^*$}.
(Note the $*$ symbol.)

\begin{definition}
\label{De:PrecisionOracle}
A \textit{precision query} takes as input an $L$-transformer $\abstfe$ (that is consistent with the set of examples $E$) and returns
\begin{enumerate}
  \item
    $\mathit{False}$ if there exists an abstract example $\langle a, c'\rangle$,
    where $a\in A$ is an abstract input, and $c'$ is a concrete
    value such that there exists a best $L$-transformer $\bestl\in \bestset$
    for which $c'\in \gamma(\abstfe(a)) \setminus \gamma(\bestl(a))$,
    and $\bestl$ is also \emph{consistent with the set of examples} $E$.
    $\langle a, c'\rangle$ is a witness that $\abst{f}{E}$ is not a best $L$-transformer;
    $\langle a, c'\rangle$ is called a \textbf{\emph{negative counterexample}}.
  \item
    $\mathit{True}$ otherwise.
\end{enumerate}
\end{definition}

Let us start by considering the case when $\bestset = \{ \bestl \}$ is a singleton set.
Given a precision query, we could try to solve our problem using a CEGIS
algorithm that uses the precision query iteratively, to synthesize
successively more precise $L$-transformers.
This (hypothetical) algorithm would maintain a set of examples $\nex$
that we want our $L$-transformer to avoid, and would issue a precision query to determine
whether the current candidate $L$-transformer is a most-precise $L$-transformer.
If not, the algorithm would add the example returned by the query to $\nex$ and repeat.
For example, starting with 
$
  \mathtt{\abst{abs}{\emptyset}(a) = [ -\infty, +\infty]},
$
\func{CheckPrecision$^*$} may generate a negative counterexample
$\B{[1,6], 10}$ that improves the precision of the $L$-transformer,
after which a new $L$-transformer is synthesized (and $\nex = \{ \B{[1,6], 10} \}$
starts to build up):
$
  \abst{\texttt{abs}}{\nex}\mathtt{(a) = [0,\ a.l+a.r]}.
$
\figref{shrinking-transformer} illustrates how this algorithm
synthesizes a more-precise $L$-transformer on each iteration.
(Green circles represent positive examples, and red stars represent negative counterexamples.)

The precision query only returns a negative counterexample when there exists
an $L$-transformer that satisfies the current set of examples $E$.
We can illustrate the definition by returning to the situation
depicted in \figref{bunny}, where there are two best $L$-transformers,
$\bestset = \{\abst{\widehat{f}}{1L}, \abst{\widehat{f}}{2L}\}$.
Suppose that $\pex=\{p_1\}, \nex=\{n1\}$, and that \func{Synthesize}
has found best $L$-transformer $\abst{\widehat{f}}{1L}$
(so we would like the algorithm to terminate).
At this point, the only other best $L$-transformer that
the precision query could find is $\abst{\widehat{f}}{2L}$.
It is true that $n_2 \in \gamma(\abst{\widehat{f}}{1L}) \setminus \gamma(\abst{\widehat{f}}{2L})$,
but if $n_2$ were added to $\nex$, \func{Synthesize} would then be blocked from
finding \emph{any} $L$-transformer consistent with $E$,
and thus a CEGIS procedure would fail (without returning any $L$-transformer).
However, a best $L$-transformer returned by the precision query must satisfy both $\pex$ and $\nex$,
and $\abst{\widehat{f}}{2L}$ does not satisfy $\nex$ (which contains $n_1$).
The requirement to satisfy $E$ prevents the precision query from returning
$\abst{\widehat{f}}{2L}$;
instead, the precision query would return $\texttt{True}$, and
CEGIS would terminate with $\abst{\widehat{f}}{1L}$.

The following lemma describes how the \func{CheckPrecision$^*$} query relates to the
problem of synthesizing a best $L$-transformer.
\begin{lemma}
Suppose that $\abst{f}{\empty}$ is a sound $L$-transformer for $f$.
$\abst{f}{\empty}$ is a best $L$-transformer for $f$ if and only if
for every set of examples $E$ with which $\abst{f}{\empty}$ is consistent,
the answer to the query \func{CheckPrecision$^*$}
with respect to $\abst{f}{\empty}$ and examples $E$ is $\mathit{True}$.
\end{lemma}
The lemma shows how a $\mathit{False}$ answer from \func{CheckPrecision$^*$}
guarantees that a more precise $L$-transformer exists in $L$,
but a positive answer on a \emph{single} set of examples $E$ does not
guarantee that the $L$-transformer $\abst{\widehat{f}}{\empty}$ is a
best $L$-transformer.

\subsubsection*{From \func{CheckPrecision$^*$} to \func{CheckPrecision}. }
Given an L-transformer $\abstfe$, the precision query \func{CheckPrecision$^*$}
returns true if $\abstfe$ is
a maximally precise overapproximation
of $\bestl$ (with respect to $\sqsubseteq_{pr}$); otherwise it returns a counterexample $\B{a,c'}$.
Even though a logical specification $\Phi_f$ for the semantics of the
concrete function $f$ is provided, where $\Phi_f$ is expressed in a
decidable logic, instantiating \func{CheckPrecision$^*$} would be
challenging because $\bestl$ is not known.

\Omit{
In the case when a logical specification is not provided, an approximation to
\func{CheckSoundness$^*$} can be built on top of
    a closed-box oracle $\mathcal{R}un(f,\cdot)$ for the concrete function
    $\llbracket{f}\rrbracket$ using
    a \emph{semi-algorithm} that searches for a counterexample
    to soundness by
    calling $\mathcal{R}un(f,c)$ on different concrete inputs $c$.
}

Instead, \func{CheckPrecision}---note $*$ symbol---uses the current
set of examples $E$ to approximate $\bestl$:
a most-precise $L$-transformer that satisfies $E$ (denoted by $\beste$) is
optimistically \textit{assumed} to be $\bestl$.
Of course, this approximation improves as more positive examples are discovered.
Furthermore, we do not need to compute $\beste$;
any $L$-transformer $\abst{h}{L} \sqsupseteq_{\textit{pr}} \beste$ (where $\abst{h}{L}$ satisfies
certain other conditions) suffices, as we explain next.

\func{CheckPrecision} attempts to discover a negative counterexample ($e^-$),
while ensuring that there exists a sound $L$-transformer $\abst{h}{L}$
that continues to satisfy both $\pex$ and $\nex$ of example set $E$,
in addition to satisfying the negative counterexample $e^-$.
Clearly, $\abst{h}{L} \sqsupseteq_{\textit{pr}} \beste$ for some $\beste$.
%
%
Given a candidate transformer $\abstfe$, \func{CheckPrecision} asserts the following conditions:
\begin{itemize}
    \item
      All examples in $\pex$ and $\nex$ are satisfied (see \eqref{satsynth}).
    \item
      There exists a feasible $L$-transformer $\abst{h}{L}$ that satisfies $E$, as well as a
      new negative counterexample $\B{a, c'} \not\in \nex$
      \[
         sat^+(\abst{h}{L}, \pex) \land sat^-(\abst{h}{L}, \nex) \land sat^-(\abst{h}{L}, \{ \B{a, c'} \}).
      \]
    \item The transformer $\abstfe$ does not satisfy the new (negative) example $\B{a, c'}$
       \[
         sat^-(\abstfe, \{ \B{a, c'} \}) = \textit{false}.
       \]
\end{itemize}

\noindent
The precision check can be expressed as follows:
\begin{align}
    \label{Eq:checkPrecisionCondition}
    \exists \abst{h}{L}, \langle a, c' \rangle. \
              sat^+(\abst{h}{L}, \pex)
        \land sat^-(\abst{h}{L}, \nex \cup \{ \langle a, c'\rangle \})
        \land \neg sat^-(\abstfe, \{ \B{a, c'} \})
\end{align}
We can now define the \func{CheckPrecision} interface:
\begin{equation}
\label{Eq:checkprecision}
  \func{CheckPrecision}(\abstfe, \pex, \nex) =
                        \begin{cases}
                             \textit{False}, \B{a,c'}   &\mbox{if \eqref{checkPrecisionCondition} is SAT}\\
                             \textit{True}, \_          &\mbox{otherwise}
                           \end{cases}
\end{equation}


\section{An Algorithm to Synthesize a Best $L$-Transformer}
\label{Se:Algorithm}

\subsection{Accommodating Competing Objectives}
\label{Se:CompetingObjectives}

The competing objectives of soundness and precision might seem to stand in
the way of designing an algorithm that can achieve both.
To address this issue, our algorithm essentially runs \emph{two} CEGIS loops---one
for soundness and one for precision.
At each step, it non-deterministically chooses to query
\func{CheckSoundness} or \func{CheckPrecision}, improving soundness or
precision, respectively.
When neither query generates any further counterexamples, the
algorithm has provably synthesized a best $L$-transformer.

Because the two CEGIS loops operate independently, interacting only
through examples, improving soundness can temporarily compromise
precision and vice-versa.
For example, the first positive counterexample could lead the
synthesizer to emit $\lambda a. \top$, allowing progress only via a
negative counterexample.
Similarly, a negative counterexample can temporarily compromise soundness.
(See \sectrefs{Algorithm}{Implementation}.)


\begin{example}
\label{Exa:AbstractAbsIllustration}
We now illustrate the principles used in the algorithm on the
$\texttt{abs}$ function, using the interval abstract domain and the
DSL from \eqref{LForIntervalDomain}.
To start, $\mathtt{\abst{{abs}}{\B{\emptyset,\emptyset}}(a) = \bot}$. 
Now suppose that \func{CheckSoundness} is used to generate a positive counterexample, say,
$\B{[5,5], 5}$.
The algorithm records this example, and attempts to synthesize an $L$-transformer that satisfies it, say,
\[
  \mathtt{ \abst{abs}{\B{ \{ \B{[5,5], 5} \}, \emptyset}} (a) = [a.l, \infty]}.
\]
Because the above $L$-transformer is still not sound, there could be
additional calls to \func{CheckSoundness} to generate new
counterexamples to soundness.
However, the algorithm can choose nondeterministically to perform a precision step,
calling \func{CheckPrecision} to generate a negative counterexample.
Suppose that \func{CheckPrecision} generates
$\B{[1,6], 10}$, after which the following $L$-transformer is synthesized:
\[
 \mathtt{ \abst{abs}{\B{ \{ \B{[5,5], 5} \}, \{ \B{[1,6], 10} \}}} (a) = [0,\ a.l+a.r]}.
\]
Eventually, the algorithms will not be able find any additional positive or negative
examples, and will return the following sound and precise $L$-transformer:
\[
  \abst{\texttt{abs}}{E} \mathtt{(a) = [max(max(0, a.l), -a.r), \ max(-a.l, a.r)]}.
\]
\end{example}

\subsection{Consistency of Positive and Negative Examples}
\label{Se:ConsistencyOfPositiveAndNegativeExamples}

We define the following interface functions:
\begin{equation*}
  \label{Eq:checkconsistency}
  \textsc{CheckConsistency}(\pex, \nex) = \exists \abstfe.\ sat^+(\abstfe, \pex) \land sat^-(\abstfe, \nex)
\end{equation*}
    
\begin{equation}
  \label{Eq:synthesize}
  \textsc{Synthesize}(\pex, \nex) =
  \begin{cases}
       \abst{f}{E}  & \textrm{if}~\exists \abstfe.\ sat^+(\abst{f}{E}, \pex) \land sat^-(\abst{f}{E}, \nex) \\
       \bot         & \mbox{otherwise}
   \end{cases}
\end{equation}

In our algorithm, \textsc{Synthesize} is only called if \textsc{CheckConsistency} returns true and is, therefore, guaranteed to return a transformer. 

\begin{wrapfigure}{R}{.45\linewidth}
    \vspace{-2.5ex}
    \centering
    \frame{\includegraphics[scale=.55]{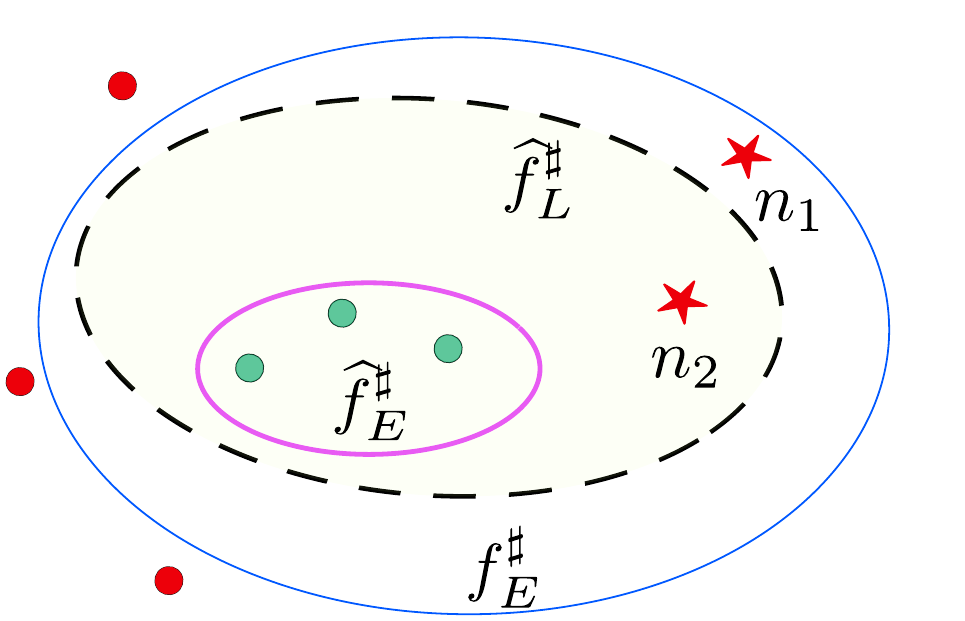}}
    \caption{\label{Fi:consistent}Negative examples: $n_1$ is outside of $\bestl$ and $n_2$ is inside $\bestl$.}
\end{wrapfigure}

Recall that whenever a new negative counterexample $e^-$ is generated for a candidate transformer $\abstfe$,
\func{CheckPrecision} uses $\abst{h}{L}$ as an over-approximation ($\sqsupseteq_{\textit{pr}}$) of
(some) $\beste$.
Because $e^-$ is excluded from $\abst{h}{L}$, $e^-$ is also excluded from some $\bestl$.
However, there are two possible cases:

\begin{enumerate}
    \item
    \label{It:NegativeCounterExampleOne}
    The negative counterexample $\B{a, c'}$ is such that
    $c' \notin \gamma(\abst{\widehat{f}}{L}(a))$ (and
    $c' \in \gamma(\abst{f}{E}(a))$).
    This case is illustrated by the red star labeled $n_1$ in \figref{consistent}, where
    $\abst{\widehat{f}}{L}$ and ${\abst{f}{E}}$ are shown as the dashed black and solid blue ovals, respectively.
    In this case, $n_1$ is inside ${\abst{f}{E}}$, but outside $\abst{\widehat{f}}{L}$.

\item
    \label{It:NegativeCounterExampleTwo}
    The negative counterexample $\B{a, c'}$ is such that
    $c' \in \gamma(\abst{\widehat{f}}{L}(a))$ (and
    $c' \in \gamma(\abst{f}{E}(a))$).
    Hence, by excluding this negative counterexample, the synthesized transformer
    $\abst{f}{E}$ remains sound w.r.t.\ the examples $E$, but can become unsound
    w.r.t.\ $\abst{\widehat{f}}{L}$.
    This case is illustrated by the red star labeled $n_2$ in \figref{consistent}:
    $n_2$ is inside both ${\abst{f}{E}}$ and $\abst{\widehat{f}}{L}$.
\end{enumerate}

If $\pex = \{ \langle[1,5], 2\rangle \}$ and $\abst{f}{E}$ is $\lambda a . [a.l, +\infty] $,
the negative example $\langle[-15,-11],2\rangle$
illustrates an example of case \ref{It:NegativeCounterExampleTwo} above, and
the transformer $\lambda a. [a.l, a.r]$ would be a possible $\abst{h}{L}$.

Because $\bestl$ is unknown, our algorithm has no means for
identifying which of the above cases a negative counterexample falls
into.
Hence, the algorithm tentatively marks the example as a negative example. 
However, such an assumption has the risk of making the set of positive
and negatives examples inconsistent:
the examples $E = \B{\pex, \nex}$ are inconsistent if there does not exist any $L$-transformer
$\abst{g}{E}$ such that $sat^+(\abst{g}{E}, \pex) \land sat^-(\abst{g}{E}, \nex)$.

\begin{figure}
	\label{Fi:checkinconsistent}
  \begin{minipage}[t]{0.45\linewidth}
    \centering
    \frame{\includegraphics[width=0.8\linewidth]{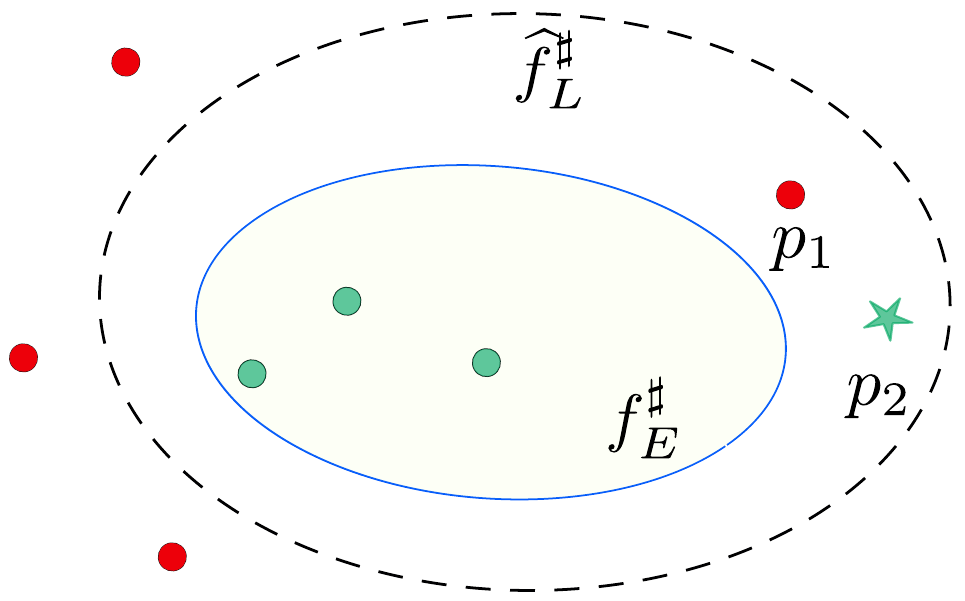}}
    \subcaption{Positive example $p_2$ creates an inconsistency w.r.t.\ $p_1$.}
    \label{Fi:inconsistent2}
  \end{minipage}
\hspace{1cm}
  \begin{minipage}[t]{0.45\linewidth}
    \centering
    \frame{\includegraphics[width=0.8\linewidth]{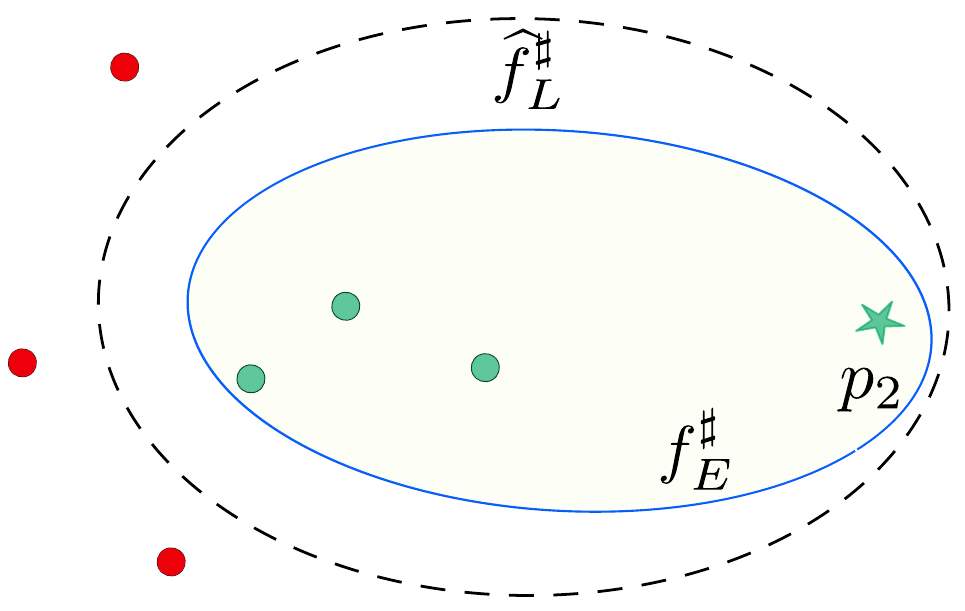}}
      \subcaption{The negative example $p_1$ is dropped.}
    \label{Fi:inconsistent3}
  \end{minipage}

  \caption{(a) Inconsistent positive and negative examples. (b) Illustration of how our algorithm resolves the inconsistency. (\textcolor{green!70!blue}{$\bigstar$}: most recent positive example)}
\end{figure}

In our example, suppose that in the next iteration the algorithm selects the soundness check,
calling \func{CheckSoundness} to generate another positive example,
say $\B{[-1,0],1}$ ($p_2$ in \figref{inconsistent2}).
Now the examples are inconsistent because there does not exist any $L$-transformer that includes
$\B{[-1,0],1}$ and excludes $\B{[-15, -11], 2}$.
At this point, {\func{CheckConsistency}} finds that the positive and negative examples are inconsistent, and hence, some negative example added by \func{CheckPrecision} must have been a positive example.

The algorithm uses Occam's razor to solve this dilemma: the goal of the synthesis step becomes
``synthesize a transformer that ignores the \textit{smallest number of negative examples}.''
In our case, by ignoring the negative example $\langle [-15, -11], 2\rangle$,
it is possible to synthesize the abstract transformer,
\begin{equation*}
  \fSharpSyn  \mathtt{(a: \intv)} : \mathtt{\intv }=  \mathtt{[max(max(0, a.l), -a.r), \ max(-a.l, a.r)]}
\end{equation*}

To reestablish the consistency of the sets of positive and negative examples,
the algorithm now drops the negative examples that it could not satisfy from $\nex$.
This scenario is illustrated in \figref{inconsistent3}, where the
negative example $p_1$ is dropped, so
that positive and negative examples are now consistent.
Hence, in addition to a procedure, \func{Synthesize},
the algorithm needs access to a stronger synthesis procedure, \func{MaxSatSynthesize}.
This procedure is similar to the standard partial MaxSat procedure~\cite{li2009maxsat}
(allowing for hard and soft constraints), but applied in the context of
program synthesis.
When the positive and negative example sets are conflicting,
\func{MaxSatSynthesize} attempts to synthesize a transformer by
dropping the smallest number of negative examples that make the query
satisfiable.

\begin{definition}
Given example set $\B{\pex, \nex}$ for which there is no $L$-transformer $\abst{g}{E}$
such that $\textsc{CheckConsistency}(\abst{g}{E}, \pex, \nex) = \textit{true}$,
\textsc{MaxSatSynthesize} returns an $L$-transformer that can be
synthesized by dropping the smallest set of negative examples:
    
    \begin{align}    
      &\textsc{MaxSatSynthesize}(\pex, \nex) = 
      & \begin{cases}
           \abst{f}{E}, D  &\textrm{if}~\exists \abst{f}{E}, D. \  sat^+(\abst{f}{E}, \pex)  \land sat^-(\abst{f}{E}, \nex \setminus D), \\&\textit{where $D$ is minimal},\\
           \bot  &\mbox{otherwise}
       \end{cases}
    \end{align}
\end{definition}


{\small
\begin{algorithm2e}[tb]
  \caption {\textsc{SynthesizeTransformer}($f$)  // $C, {(A, \sqsubseteq, \bot)}, \gamma, L$\label{Alg:main}}

  $\abst{f}{E} \leftarrow \lambda a: \bot$\\
  $isSound, isPrecise \gets \textit{False}$\\
  $\pex, \nex \gets \func{InitializeExamples()}$ \label{Alg:main:sampling}

  \While{$\lnot \textit{isSound} \lor \lnot \textit{isPrecise}$} {      \label{Alg:main:outerwhile}
    \If{$\lnot isSound \land \lnot isPrecise$} {
      \If{$\func{CheckConsistency}(\pex, \nex)$} {
	\label{Alg:main:isconsistent}
	$\abst{f}{E} \gets \func{Synthesize}( \pex, \nex)$ \label{Alg:main:gentransformer}
      }
      \Else {
	$\abst{f}{E}, \delta \gets \func{MaxSatSynthesize}( \pex, \nex)$\\
    \eIf{$\abst{f}{E} \neq \bot$}{
        $\nex \gets \nex \setminus \delta$ \label{Li:DropNegExample}
    }{
        \Return $\bot$
    }
      }
    }
    \If{$*$} { \label{Alg:main::if-nondet}
      { \label{Alg:main:sound}
	$\textit{isSound}, e \gets \func{CheckSoundness}(\abst{f}{E}, f)$  \label{Alg:main:CallCheckSoundness} \\
        \If{$\lnot \textit{isSound}$}{
          {$\textit{isPrecise} \gets \textit{False}$}\\
	  $\pex \gets \pex \cup \{e\}$ \label{Alg:main:AdjustEPlus} \label{Li:AdjustEPlus}
        }
      }
    }

    \Else {
      { \label{Alg:main:precise}
	$\textit{isPrecise}, e \gets \func{CheckPrecision}(\abst{f}{E}, \pex, \nex)$ \label{Alg:main:CallCheckPrecision} \\
        \If{$\lnot isPrecise$}{
          {$\textit{isSound} \gets \textit{False}$}\\
	  $\nex \gets \nex \cup \{e\}$ \label{Alg:main:AdjustEMinus} \label{Li:AdjustEMinus}
        }
      }
    }

  }
  \Return { $\abst{f}{E}$ }
\end{algorithm2e}
}

\subsection{Putting It All Together}
\label{Se:PuttingItAllTogether}
The algorithm to synthesize a best $L$-transformer is given as Alg.~\ref{Alg:main}.
In essence, it runs two CEGIS loops, attempting to meet the dual goals
of soundness and precision.
The two CEGIS loops interact with each other via the set of positive
and negative counterexamples that they generate, and the algorithm can
terminate only when both loops have attained their objectives.
The algorithm starts off with a trivial transformer that always
returns the abstract bottom, and (potentially empty) example sets
$\pex$ and $\nex$ (line \ref{Alg:main:sampling}).
While the synthesized transformer is either unsound or imprecise (line \ref{Alg:main:outerwhile}),
the algorithm synthesizes a transformer
(line~\ref{Alg:main:gentransformer}) that is consistent with the
current set of positive ($\pex$) and negative ($\nex$) examples.
Then, it non-deterministically chooses to perform either a soundness
check (line \ref{Alg:main:sound}) or a precision check (line
\ref{Alg:main:precise}), expanding its set of examples (positive or
negative, respectively) in each case if the check fails.


As discussed in \sectref{ConsistencyOfPositiveAndNegativeExamples}, it is possible that the positive and negative examples from \textsc{CheckSoundness} and \textsc{CheckPrecision} become inconsistent. 
If the examples are inconsistent (i.e., \func{CheckConsistency} fails in line~\ref{Alg:main:isconsistent}),
the algorithm invokes \func{MaxSatSynthesize}
to synthesize a transformer that satisfies all
positive examples and a maximal set of negative examples, returning
the unsatisfied negative examples in $\delta$.
The examples are adjusted to reinstate consistency (dropping the unsatisfied
negative examples), and the loop continues.
If \textsc{MaxSatSynthesize} fails to synthesize a transformer, then
it must be the case that language $L$ is inadequate to express a
valid abstraction of $f$, and the algorithm terminates with $\bot$.

The invariant of Alg.~\ref{Alg:main} is that, on each iteration, at line
\ref{Alg:main::if-nondet} a transformer $\abst{f}{E}$
has been computed that is (i)~expressible in $L$, and
(ii)~consistent with respect to $\B{\pex,\nex}$.
Then, in lines \ref{Alg:main:sound}--\ref{Alg:main:AdjustEMinus},
soundness or precision is checked.
When both hold, $\abst{f}{E}$ is a suitable $L$-transformer,
and Alg.~\ref{Alg:main} terminates.

Alg.~\ref{Alg:main} is stated as a non-deterministic procedure, as a way to separate mechanism from policy.
However, any fair scheduler can be used to resolve the non-determinism---e.g., alternating between
\func{CheckSoundness} and \func{CheckPrecision} in successive iterations.
(We describe the scheduler that we use in our implementation in \sectref{Implementation}.)

\begin{theorem}\label{The:Termination}
  If Alg.~\ref{Alg:main} terminates with a valid $\abst{f}{E}$ (i.e.,
  does not terminate with $\bot$), then $\abst{f}{E}$ is a best
  $L$-transformer for the concrete function $f$.
\end{theorem}

\begin{proof}
For the algorithm to return a valid $L$-transformer,
both \textsc{CheckSoundness}~(\eqref{checksoundness}) and \textsc{CheckPrecision}~(\eqref{checkprecision})
must have returned true (lines \ref{Alg:main:outerwhile}, \ref{Alg:main:CallCheckSoundness}, and \ref{Alg:main:CallCheckPrecision}).
  
Passing the test of \textsc{CheckSoundness} guarantees that the $L$-transformer is sound.
\textsc{CheckPrecision} attempts to find a more precise $L$-transformer
than $\abst{f}{E}$ that still satisfies the current set of positive
examples;
the inability of \textsc{CheckPrecision} to find such an $L$-transformer
proves that $\abst{f}{E}$ must also be
a best $L$-transformer.

\end{proof}

\begin{theorem}\label{The:FiniteDSL}
Let $L$ be the DSL whose syntax and semantics has been supplied by the user.
If $L$ is a finite language (i.e., the grammar for $L$ generates only a finite number of terms),
and the non-deterministic choice at line~\ref{Alg:main:sound} is resolved by a fair scheduler,
then Alg.~\ref{Alg:main} always terminates.

\end{theorem}

\begin{proof}
We assume that the executions of Alg.~\ref{Alg:main} are fair: i.e.,
the non-deterministic scheduler chooses
each of
lines \ref{Alg:main:CallCheckSoundness} and
\ref{Alg:main:CallCheckPrecision} infinitely often.
Consider a ``normalized'' execution trace, in which all calls to
\textsc{CheckSoundness} on a
sound $L$-transformer
and \textsc{CheckPrecision} on a
precise $L$-transformer
are filtered out.
That is, each remaining call on \textsc{CheckSoundness} adds to
$\pex$, and each remaining call on \textsc{CheckPrecision} adds to
$\nex$.

Consider the normalized sub-trace $t$ between any two consecutive
calls to \textsc{CheckSoundness}.
Note that by the assumption of fairness, only a finite number of calls
on \textsc{CheckSoundness} and \textsc{CheckPrecision} could have been
filtered out of $t$.
Now consider the number of calls to \textsc{CheckPrecision} and
\textsc{MaxSatSynthesize} in $t$.
\begin{itemize}
  \item
    There can be at most one call on \textsc{MaxSatSynthesize}.
    The reason is that a call on \textsc{MaxSatSynthesize} establishes
    consistency among $\pex$ and $\nex$, and subsequent calls to
    \textsc{CheckPrecision} can only generate negative counterexamples that
    are consistent with the current set $\pex$.
    (That is, in the suffix of $t$ after the call on
    \textsc{MaxSatSynthesize}, $\pex$ and $\nex$ always pass
    \textsc{CheckConsistency}.)
  \item
    In effect, each call to \textsc{CheckPrecision}
    removes at least one $L$-transformer from consideration
    (by generating a new negative counterexample that is added to $\nex$).
    Thus, when DSL $L$ is finite, there can only be a finite number of calls
    on \textsc{CheckPrecision} in $t$.
\end{itemize}

Now consider the full normalized trace.
Each call on \textsc{CheckSoundness} adds a new positive example to
$\pex$, and thus removes 
at least one $L$-transformer
from consideration during a subsequent call on \textsc{Synthesize}.
Consequently, there can only be a finite number of calls on
\textsc{CheckSoundness} in the normalized trace, and hence every
execution of Alg.~\ref{Alg:main} with a finite DSL $L$ must terminate.
\end{proof}

\begin{corollary}\label{Cor:FiniteDSLGiveBestLTransformer}
Assuming the same premises as \theoref{FiniteDSL},
if Alg.~\ref{Alg:main} terminates with a valid $\abst{f}{E}$ (i.e.,
does not terminate with $\bot$), then $\abst{f}{E}$ is a best
$L$-transformer for the concrete function $f$.
\end{corollary}

\section{Implementation}
\label{Se:Implementation}

We implemented our framework in a tool, called \tool.
\tool is written in Python and uses the Sketch synthesizer (v. 1.7.5)~\cite{Lezama13} as a subroutine.
\tool's inputs are
\begin{enumerate}
  \item
    The type of the elements in the concrete domain, e.g., integers.
  \item
    A logical specification 
    $\Phi_f$ of the semantics of concrete function $f$.
  \item 
    An implementation, written in Sketch, that specifies the
    partial order ($\sqsubseteq$) in the abstract domain,
    along with an implementation of
    \texttt{gammaCheck($c, a$)}, which, given $c \in C$ and $a\in A$,
    checks whether $c \in \gamma(a)$.
  \item The definition of the DSL, written in the Sketch language.
  \item Optionally, a set of initial positive and negative examples.
\end{enumerate}

Sketch is built on top of the C programming language, and allows one
to write programs with holes, assertions, and a
minimization
objective over an integer expression.
The goal of the Sketch solver is to find integer values for the holes
that cause all assertions to hold, while
minimizing
the value of the given objective.
As discussed below, \tool uses these features to implement the primitives
used in Alg.~\ref{Alg:main}.

\begin{figure}
\begin{subfigure}[b]{.5\linewidth}
  \begin{tabular}{c}
  \begin{lstlisting}
generator interval genT(interval inv) {
   int t = ??;
   if(t == 0) return inv;    $\label{Li:ProductionArg}$
   if(t == 1) return [0, 0]; $\label{Li:ProductionZero}$
   $\ldots$
   interval x = genT(inv);
   interval y = genT(inv);
   $\ldots$
   if(t == 2) return addInterval(x, y);       $\label{Li:ProductionAdd}$
   if(t == 3) return subtractInterval(x, y);  $\label{Li:ProductionSub}$
   $\ldots$
}
  \end{lstlisting}
  \end{tabular}
  \caption{        
           A snippet of a Sketch generator for the DSL shown in \eqref{LForIntervalDomain}.
           In essence, \linerefspp{ProductionArg}{ProductionZero}{ProductionAdd}{ProductionSub}
           are ``productions'' in a grammar with a supplied semantics
           that gives an interpretation over interval arithmetic. \label{Fi:sketchSnippet}
          }
\end{subfigure}
\hfill
\begin{subfigure}[b]{.45\linewidth}
  \begin{lstlisting}[mathescape=true]
interval $\texttt{fun}^\sharp$(interval inv){
   if(isBot(inv)){
      return $\bot$;
   }
   else{
      return genT(inv);
   }
}
  \end{lstlisting}
  \caption{Representative function for the interval-domain transformer, which returns $\bot$ if the input is also $\bot$; otherwise it returns the output of generator \texttt{genT}. \label{Fi:callSketchGen}}
\end{subfigure}
\caption{Template for the abstract transformers using a Sketch generator.\label{Fi:templateSketchGen}}
\end{figure}

One specifies a DSL $L$ in Sketch using generators.
Generators are special functions---possibly recursive---that contain
holes, and allow one to build complex programs via recursion.
Sketch allows one to set a bound on how deep the recursion can
be---thereby bounding the size of a generated program.
Generators contain holes that allow the Sketch solver to pick
productions to build a program that satisfies a given specification.
\figref{sketchSnippet} shows how the DSL from
\eqref{LForIntervalDomain} can be specified using a generator.
In such an encoding, a generator does two things simultaneously: 
(1) it uses the hole value to select some production, and
(2) it ``executes'' the code corresponding to the production.
\figref{sketchSnippet} is an example of a Sketch idiom:
it sidesteps generating an intermediate object (e.g., an abstract
syntax tree) that then needs to be executed by an interpreter
to obtain the semantics.
\twrchanged{
\figref{callSketchGen} shows the template for the abstract transformers for the interval domain.
It takes as input an interval, and returns $\bot$ in case the input is $\bot$;
otherwise, it returns the output of the call on the generator. 
}

\twrchanged{
A generator for an interval-to-interval transformer has a signature of the following form:
}
\[
  \texttt{generator interval genT(interval in)\{...\} }
\]
Sketch allows imposing constraints on the programs
produced by a generator.
For instance, one can provide a set of input/output pairs,
and assert that the program exhibits those behaviors, e.g.,
\[
  \mathtt{assert\ fun^\sharp([1,2])==[2,3]\ \&\&\ ...\ \&\&\ fun^\sharp([1,3])==[2,4]}
\]
The solver
will then compute a realization of \texttt{genT} (i.e., an $L$-transformer) that correctly
matches the given examples.

Because our algorithm may require synthesizing transformers that are
correct on ``most'' of a set of examples, we can use Sketch to count
how many examples are satisfied, and then maximize that value.
Although there is no maximize primitive, it can be simulated using
the \texttt{minimize} construct:

\[
\begin{array}{l}
  \texttt{c = NumExamples;}
  \\
  \mathtt{if(fun^\sharp([1,2])==[2,3])\ c=c-1;\ \ldots}
  \\
  \texttt{minimize(c);}
\end{array}
\]

In \sectref{Experiments}, we show some of the grammars for the DSLs
used in our experiments, e.g.,
\eqrefspp{DSLForContainsCI}{DSLForTrimCI}{TrimPSDSL}{LForIntervalDomainMul},
each of which is encoded using the idiom described above.
Because the grammars are recursive, their languages are of infinite cardinality;
however, Sketch imposes a (user-controllable) unrolling bound on generators
that are recursive.
\corref{FiniteDSLGiveBestLTransformer} holds for all of our experiments with
\tool---i.e., for a concrete function $f$, if Alg.~\ref{Alg:main} terminates
with something other than $\bot$, then the function $\abst{f}{E}$ obtained
is a best $L$-transformer for $f$---however, $L$ is the finite language of
programs that fall within the unrolling bound, not the full
infinite-cardinality language of the grammar.
\twrchanged{
The current implementation indicates the existence (yes/no) of a bug in a manually written transformer.
To locate a bug, a user must currently inspect the faulty transformer and the one synthesized by \tool manually.
}

\subsubsubsection{\func{CheckSoundness}.}
    Given a logical specification $\Phi_f$ of concrete function $f$, we use
    Sketch as a satisfiability solver.
    For a given candidate $L$-transformer $\abstf$, the \texttt{gammaCheck} primitive is used to specify that
    Sketch should try to find an input $c \in \gamma(a)$, such that $\Phi_{f}(c) \notin \gamma(\abstf(a))$.
    If successful, $\B{a,\Phi_{f}(c)}$ is returned as a
    positive counterexample.

\FuzzOut{When a logical specification is not available, the soundness oracle is approximated using AFL to find positive counterexamples (i.e., by running the concrete function on a number of examples).  Given a candidate transformer $\abstf$, and access to a closed-box oracle $\llbracket f \rrbracket$ for the concrete semantics of $f$, the fuzzer repeatedly runs $\llbracket f \rrbracket$ on different inputs.  Here, $f$ is augmented with an assertion that fails whenever the fuzzer generates an input $c \in \gamma(a)$, for some $a\in A$, such that $\llbracket f(c) \rrbracket \notin \gamma(\abstf(a))$.  If an assertion failure is found, $\B{a,\llbracket f(c) \rrbracket}$ is returned as a positive counterexample.  If the fuzzer times out, we assume that $\abstf$ is sound.}

\FuzzOut{Checking soundness is an undecidable problem, and our implementation
is not guaranteed to synthesize sound transformers;
however, we manually inspected each transformer synthesized in our benchmarks, and in every case,
we found that the transformer synthesized by {\tool} was sound.}

\subsubsubsection{\func{CheckPrecision}, \func{CheckConsistency}, \func{Synthesize}, and \func{MaxSatSynthesize}.}
Although in Alg.~\ref{Alg:main} \func{CheckPrecision}, \func{CheckConsistency}, and
\func{Synthesize} are shown as separate procedure calls, {\tool} implements all
three calls via a single invocation of Sketch
(which finds a new $L$-transformer
that is more precise than the given one, as well as a witness example).
\func{MaxSatSynthesize} uses the \texttt{minimize} construct from Sketch, as described above.

\subsection{Designing a DSL}
\label{Se:DesigningADSL}

As in all synthesis tasks, the design of the DSL is important.
DSLs for expressing abstract transformers essentially involve a combination of
\begin{enumerate}
  \item
    \label{It:OperationsOnConcreteValues}
    Primitives that operate on concrete values (e.g., addition,
    subtraction, etc.\ for numeric values; isSubset, size, containsSpace,
    etc.\ for strings).
    The reason is that concrete values may appear as components of
    abstract values, such as endpoints of an interval or a member of a
    string set.
\twrchanged{
  \item
    \label{It:DeconstructAbstractValues}
    Operations to deconstruct an abstract value (e.g., to access one of the two endpoints of an interval).
  \item
    \label{It:ConstructAbstractValues}
    Operations to construct an abstract value (e.g., to pair the two endpoints of an interval).
}
  \item
    \label{It:BooleanConnectives}
     Boolean connectives.
  \item
    \label{It:ControlFlowConstructs}
     Control-flow constructs.
\end{enumerate}

\twrchanged{
Consider the DSL shown in \eqref{LForIntervalDomain}.
The production $\textit{Transformer}~::=~\lambda \texttt{a} . [E, E]$ is an example of
\itemref{ConstructAbstractValues}:
an interval is created from the values of two expressions.
Nonterminal $E$ derives expressions for use in a transformer.
The productions $E~::=~\texttt{a.l} \mid \texttt{a.r}$ are examples of \itemref{DeconstructAbstractValues}:
here \texttt{a} refers to the $\lambda$-bound variable in $\lambda \texttt{a} . [E, E]$, and
\texttt{a.l} (\texttt{a.r}) selects the left (right) end of interval \texttt{a}.
The remaining productions of nonterminal $E$ are examples of \itemref{OperationsOnConcreteValues}:
they allow synthesizing arithmetic operations, e.g., sums, differences, products, etc.
This DSL is a natural fit for the problem of creating an interval
transformer for the absolute-value function.
}

\twrchanged{
The key point in our work is that we have a DSL $L$, and (provably)
obtain one of the best results possible in $L$.
For a given DSL grammar $G$, one has different languages $L_1$, $L_2$, etc.
according to the grammar depth that one chooses to use.
Thus, if you increase the grammar depth from $k$ to $k+1$,
you would then obtain a best $L_{k+1}$-transformer, which can be more
precise than a best $L_k$-transformer.
In the experiments described in \sectref{Experiments}, which involved
synthesizing 57 transformers (for 15 operations and 8 abstract domains:
$57 = 6 \times 5 + 9 \times 3$; see \tableref{OperationList}),
we found that most transformers could be synthesized
with reasonably simple grammars and small grammar depths.
In our experiments, we used grammar depth 3 for almost all of the
transformer-synthesis tasks.
We chose 3 because we found that we were able to synthesize almost all
transformers in a reasonable time ($<$600 seconds), and the transformers
obtained were equal to or better than (i.e., were sound) the transformers
used in {\safestr} and the tool of \citeauthor{APLAS:WrInterval12}.
}

\twrchanged{
In \sectref{ExperienceWithDesigningDSLs}, we provide additional
discussion about the design of DSLs, using example DSLs from
\sectrefs{casestudy2}{casestudy1}.
\sectref{ExperienceWithDesigningDSLs} also discusses how changing the DSL
can lead to a different (more-precise or less-precise) transformer and/or
impact the synthesis time.
}

\subsection{Refinements to Alg.~\ref{Alg:main}}
In \sectref{PuttingItAllTogether}, we stated Alg.~\ref{Alg:main} as a non-deterministic procedure,
as a way to separate mechanism from policy.
However, any fair scheduler can be used to resolve the
non-determinism (e.g., alternating between \func{CheckSoundness} and
\func{CheckPrecision} in successive iterations).
We found that a simple deterministic scheduler that prioritizes
\func{CheckPrecision} over \func{CheckSoundness} works well in
practice.
In Alg.~\ref{Alg:main}, if both the $\mathit{isSound}$
and $\mathit{isPrecise}$ flags are $\mathit{False}$,
our algorithm chooses the branch that calls \func{CheckPrecision}.
Fairness is ensured by forcing a call on \func{CheckSoundness} after
$k$ consecutive calls on \func{CheckPrecision} have been performed.
Our implementation uses this strategy (with $k=50$).




\section{Evaluation}
\label{Se:Experiments}

\begin{wraptable}{R}{.6\linewidth}
    \setlength{\tabcolsep}{3pt}
    \caption{List of abstract operations synthesized by \tool for the String and Fixed-Bitwidth Interval domains.\label{Ta:OperationList}}
{\small
    \begin{tabular}{c l c}
        \hline        
        \multirow{2}{*}{\makecell{Domain \\ Type}} & \multirow{2}{*}{\makecell{Abstract  Domains}} & \multirow{2}{*}{\makecell{Operations}} \\
        &&\\
        \hline \hline
        \multirow{5}{*}{String}  & Constant String ($\csdom$)     & \multirow{5}{*}{\makecell{$\texttt{charAt}^\sharp$, $\texttt{concat}^\sharp$, \\ $\texttt{contains}^\sharp$, \\ $\texttt{toLower}^\sharp$, $\texttt{toUpper}^\sharp$, \\ {$\texttt{trim}^\sharp$}  }}    \\
                                 & String Set (size $k$) ($\ssk$) &    \\
                                 & Char Inclusion ($\charin$)&  \\ 
                                 & Prefix-Suffix ($\psdom$)       &  \\ 
                                 & String Hash ($\mathcal{SH}$)   &   \\ 

        \hline
        \multirow{3}{*}{\makecell{Fixed\\Bitwidth\\Interval}}   & Unsigned-Int ($\uintv$)  &  \multirow{3}{*}{\makecell{$\texttt{add}^\sharp$, $\texttt{sub}^\sharp$, $\texttt{mul}^\sharp$,  $\texttt{and}^\sharp$, $\texttt{or}^\sharp$, \\ $\texttt{xor}^\sharp$, $\texttt{shl}^\sharp$, $\texttt{ashr}^\sharp$, $\texttt{lshr}^\sharp$}}  \\ 
                                   &Signed-Int ($\uintv$) &   \\ 
                                   & Wrapped ($\wintv$) &   \\

         \hline
    \end{tabular}
}
\end{wraptable}

We performed two studies with \tool.

\noindent
{\bf Case Study 1 (\sectref{casestudy2}):}
We used \tool to synthesize abstract transformers for string
operations using the multiple string abstract domains employed in
\safestr~\citep{TACAS:SAFEstr17}.
We compared the synthesized transformers with the hand-crafted
transformers implemented in \safestr.

\noindent
{\bf Case Study 2 (\sectref{casestudy1}):}
We used \tool to synthesize abstract transformers for simple
mathematical operations using three kinds of interval abstract domains
defined by \citet{APLAS:WrInterval12}.
We compared the synthesized transformers with those implemented by
\citeauthor{APLAS:WrInterval12}

See \tableref{OperationList} for the abstract domains and operations
used in these studies.
The two studies were designed to shed light on the following research questions:
\smallskip
\noindent
\begin{mdframed}[innerleftmargin = 3pt, innerrightmargin = 3pt, skipbelow=-0.25em]
  \textbf{[RQ1]:}
  How long does it take \tool to synthesize best $L$-transformers?
  \FuzzOut{
  \newline
					\textbf{[RQ2]:}
  Does the effectiveness of tool change based on whether one has a logical specification $\Phi_{\textit{op}}$ of $\llbracket \textit{op} \rrbracket$ or only closed-box access $\mathcal{R}un(\textit{op},\cdot)$
				to $\llbracket \textit{op} \rrbracket$?}
  \newline
  \textbf{[RQ2]:}
  How do the abstract transformers synthesized by \tool compare to manually written ones?
\end{mdframed}
We ran all experiments on an Intel(R) Xeon(R) 2.00GHz E5-2620 CPU with 32GB RAM, running Ubuntu 16.04.
Each reported time is the median of three runs of \tool.
We used an unrolling depth of 3 for the DSLs used in our experiments.
We used a timeout value of 600s per call on Sketch.

\subsection{Case Study 1: Transformers for the String Domains in \safestr}
\label{Se:casestudy2}

\safestr is a state-of-art static analyzer for programs involving
complex string operations.
In this study, we used \tool to create abstract transformers over the
abstract domains used in \safestr.

\subsubsection{String Abstract Domains}
\label{Se:StringDomains}

Our study considered the five string abstract domains summarized below,
which are all used in \safestr.

\subsubsubsection{String Set ($\ssk$)~\citep{CC:Magnus14}.}
This domain can abstract a finite set of strings precisely, as long as the size of the set does not exceed $k$.
An element of this domain (with the exception of $\top_{\ssk}$) is a set of constant strings of size up to $k$---i.e., $\ssk = \{\top_{\ssk}\} \cup \{S \mid   S \subseteq {\Sigma^*} \land |S|\le k\}$, where $\Sigma$ is the set of all characters, and $\bot_{\ssk}$ is the empty set.
Let $S$ be a set of strings, then the abstraction function is defined as $\alpha_{\ssk}(S) = S$ if $|S| \le k$ and $\top_{\ssk}$ otherwise. The lattice operations, $\sqcup_{\ssk}$ (join) and $\sqsubseteq_{\ssk}$ (partial order) are defined in terms of
set $\cup$ and $\subseteq$, respectively. If a set exceeds size $k$, its abstraction is $\top_{\ssk}$.

\subsubsubsection{Constant String ($\csdom$)~\citep{CC:Magnus14}.}
This domain can abstract precisely up to one concrete string;
it is a special case of the previous domain---i.e., $\csdom = \mathcal{SS}_1$.

\subsubsubsection{Character Inclusion ($\charin$) ~\citep{TACAS:SAFEstr17}.}
An element of this domain is a pair of two sets of characters, $[L, U]$.
The set $L$ (resp.\ $U$) denotes what characters a string must (resp.\ may) contain
to be in the concretization of this abstract element.
We will sometimes refer to $L$ and $U$ as \textit{must} and \textit{may} sets, respectively, in our discussion.
Formally, $\charin = \{\bot_{\charin}\} \cup \{\left[L,U\right] \mid   L, U \subseteq \Sigma, L \subseteq U\}$.
Given a string $s\in\Sigma^*$, let $char(s)$ denote the set of characters in $s$.
The abstraction function is defined as
$\alpha_\charin(\{s_1,\ldots,s_n\})=[\bigcap_i char(s_i),\bigcup_i char(s_i)]$, and
the concretization function is then defined as
$\gamma_\charin([L,R])=\{s\mid   L\subseteq char(s)\subseteq R \}$.
The partial-order relation is defined as
$[L_1, U_1] \sqsubseteq_\charin [L_2, U_2] \Leftrightarrow (L_1 \subseteq L_2 \land U_1 \subseteq U_2)$,
and the $\sqcup_{\charin}$ (join) operation is defined
as $[L_1, U_1] \sqcup_\charin [L_2, U_2] = [L_1 \cap L_2, U_1 \cup U_2]$.

\subsubsubsection{Prefix-Suffix ($\psdom$)~\citep{TACAS:SAFEstr17}.}
An element of this domain is
a pair consisting of two strings $\B{\pre,\suf}$
(we use a different pair notation to distinguish from the previous domain),
where $\pre \in \Sigma^*$ is the longest common prefix ($\lcp$) and
$\suf \in \Sigma^*$ is the longest common suffix ($\lcs$) for the corresponding set of strings.
The abstraction function is defined as
$\alpha_\psdom(S)= \B{\lcp(S), \lcs(S)}$, and
the concretization function is defined as
$\gamma_\psdom(\B{\pre,\suf}) = \{s\mid   \exists {s_1 \in \Sigma^*},\exists {s_2 \in \Sigma^*}.\ s{=}\pre.s_1\wedge s{=}s_2.\suf \}$.
The partial order is defined as $\B{\pre_1, \suf_1} \sqsubseteq _\psdom \B{\pre_2, \suf_2} \Leftrightarrow  \lcp(\{\pre_1,\pre_2\}) = \pre_2 \land \lcs(\{\suf_1, \suf_2\}) = \suf_2$,
and the join operation is defined as
$\B{\pre_1, \suf_1} \sqcup_\psdom \B{\pre_2, \suf_2} = \B{\lcp(\{\pre_1,\pre_2\}), \lcs(\{\suf_1, \suf_2\}) }$.

\subsubsubsection{String Hash ($\mathcal{SH}$)~\citep{CC:Magnus14}.}
This domain uses a hash function $h : \Sigma^* \rightarrow U$,
which takes the sum of the character codes in a string, and maps it to an element
in a fixed-size universe $U = \{0,\dots,b-1\}$.
The concrete implementation of \safestr uses the function
$h(s) = (\Sigma_{c \in char(s)}I(c))~mod~b$,
where $I:\Sigma \rightarrow \mathbb{N}$, is a mapping from characters to an integer value.
An element of the abstract domain $\mathcal{SH}$ is
a set $H\subseteq U$ denoting the hash values of the strings being tracked.
Let $S$ be a set of strings, then the abstraction function is defined as
$\alpha_\mathcal{SH}(S) = \{ h(s) \mid s \in S \}$.
The concretization function is defined as $\gamma_{\mathcal{SH}} (H) = \{s \in \Sigma^* \mid   h(s) \in H\}$.
The partial order ($\sqsubseteq_{\mathcal{SH}}$) and join ($\sqcup_{\mathcal{SH}}$)
are defined as $\subseteq$ and $\cup$ on sets, respectively.

\subsubsection{Abstract Transformers for String Operations}
\forOOPSLA{
	\begin{wraptable}{R}{.5\linewidth}
	}
	\forARXIV{
		\begin{table}
		}
		
		\caption{Time, in seconds, to synthesize abstract transformers for string functions.
			The shaded cell indicates that we used a sketch of the transformer in this case, and {\tool} only synthesizes the \textit{holes} in the sketch.}
		\label{Ta:strdom-synthesis}
		\centering
		\footnotesize
		\begin{tabular}{l|Hcrrrrr} 
			$f$ & mode & $\csdom$ & $\ssk$ & $\charin$ & $\psdom$ & $\mathcal{SH}$\\
			\hline    \hline
			{\texttt{charAt}}  & $\Phi$& 18.29 & 3.94 & 24.91 & 5.94 & 3.76 \\
			\hline
			{\texttt{concat}} & $\Phi$ & 99.05 & 9.57 & 1,983.83 & 8.92 & \cellcolor{gray!25!white} 609.30 \\
			\hline
			{\texttt{contains}} 
			& $\Phi$  & 132.06 &  78.42 & 1,804.69 & 9.13 & 10.39 \\
			
			\hline
			{\texttt{toLower}} 
			& $\Phi$ & 11.26 & 11.74 &  381.65 & 6.91  & 8.44 \\
			\hline
			{\texttt{toUpper}} 
			& $\Phi$ & 9.77 & 12.18 &  735.13 & 5.85 & 3.73 \\
			
			\hline
			{\texttt{trim}}  
			& $\Phi$ & 4.31 & 16.35 & 641.53  & 8.52 & 8.29\\
			
			\hline
		\end{tabular}
		
		\forARXIV{
		\end{table}
	}
	\forOOPSLA{
	\end{wraptable}
}

Our study involved six string-manipulation operations:
\texttt{concat}, \texttt{contains}, \texttt{charAt}, \texttt{toLower}, \texttt{toUpper}, and \texttt{trim}.
For each domain from \sectref{StringDomains},
we used \tool to synthesize abstract transformers for the six functions.
For each concrete function and abstract domain, we specified a particular DSL $L$,
and then ran \tool to synthesize a best $L$-transformer.
The time taken by \tool to synthesize an $L$-transformer across all
experiments varies between 3.73s and 1,983.83s;
see \tableref{strdom-synthesis}.
\forARXIV{
A detailed log of our experiments on synthesizing abstract transformers for abstract string
domains is given in \tableref{StringDomLogicalLog}.
The table headers should be interpreted as follows:
\begin{itemize}
	\item \textbf{Domain}: Domain name
	\item $\mathbf{f}$: Concrete function name
	\item \textbf{\#SO}: Number of times a soundness query is performed
	\item \textbf{\#PO}: Number of times a precision query is performed
	\item \textbf{\#PosEx}: Number of positive examples 
	\item \textbf{\#NegEx}: Number of negative examples 
	\item \textbf{\#MaxSat}: Number of times \textsc{MaxSatSynthesis} is invoked
	\item \textbf{\#Dropped}: Total number of examples dropped during \textsc{MaxSatSynthesis}
	\item \textbf{SO Time (s)}: Total time taken in seconds by soundness queries
	\item \textbf{PO Time (s)}: Total time taken in seconds by precision queries
	\item \textbf{Time (s)}: Total time taken in seconds
\end{itemize}

\begin{table*}[t]
	\small
	\caption{Table showing detailed results for the string-domain experiments.\label{Ta:StringDomLogicalLog}}
	\begin{tabular}{|r|r|r|r|r|r|r|r|r|r|}
		\hline
		$f$ &\#SO&\#PO&\#PosEx&\#NegEx&\#MaxSat&\#Dropped&SO Time (s)&PO Time (s)&Time (s)\\ \hline \hline
		
		\multicolumn{10}{l|}{\cellcolor{black}\textcolor{white}{\textit{\textbf{$\mathcal{CS}$}}}}\\
		
		\texttt{charAt}		& 2 & 4 & 15 & 12 & 0 & 0 & 1.75 & 16.29 & 18.29\\ \hline
		\texttt{concat}		&2 & 16 & 19 & 16 & 0 & 0 & 2.03 & 96.26 & 99.05\\ \hline
		\texttt{contains}	&2 & 28 & 15 & 29 & 0 & 0 & 1.84 & 129.6 & 132.06\\ \hline
		\texttt{toLower}	&2 &6&7&12&0&0&1.74&9.19&11.26\\ \hline
		\texttt{toUpper}	&2 & 5 & 7 & 11 & 0 & 0 & 1.77 & 7.70 & 9.77\\ \hline
		\texttt{trim}		&2 & 2 & 11 & 0 & 0 & 0 & 1.81 & 2.40 & 4.31\\ \hline
		
		\multicolumn{10}{l|}{\cellcolor{black}\textcolor{white}{\textit{\textbf{$\mathcal{SS}_k$}}}}\\
		
		\texttt{charAt}		& 2 & 2 & 5 & 0 & 0 & 0 &1.89 & 1.90 & 3.94 \\ \hline
		\texttt{concat}		& 2 & 6 & 5 & 2 & 1 & 1 & 0.93 & 6.80 & 9.57 \\ \hline
		\texttt{contains} 	& 3 & 10 & 8 & 4 & 1 & 1 & 12.55 & 63.16 & 78.42 \\ \hline
		\texttt{toLower}	& 2 & 6 & 3 & 2 & 0 & 0 & 2.19 & 7.65 &11.74 \\ \hline
		\texttt{toUpper}	& 2 & 6 & 3 & 2 & 0 & 0 & 2.16 & 8.00 & 12.18 \\ \hline
		\texttt{trim}		& 2 & 6 & 3 & 2 & 0 & 0 & 1.93 & 12.74 & 16.35\\ \hline
		
		\multicolumn{10}{l|}{\cellcolor{black}\textcolor{white}{\textit{\textbf{$\mathcal{CI}$}}}}\\
		
		\texttt{charAt}		& 3 & 4 & 14 & 6 & 0 &0 & 5.46 & 13.25 & 24.91 \\ \hline
		\texttt{concat}		& 3 & 8 & 4 & 8 & 0 & 0 & 602.53 & 1,380.92 & 1,983.83\\ \hline
		\texttt{contains}   & 5 & 25 & 15 & 17 & 2 & 2 & 607.40 &  1,189.36 & 1,804.69 \\ \hline
		\texttt{toLower}	& 2 & 3 & 7 & 1 & 0 & 0 & 202.25 & 179.28 & 381.65\\ \hline
		\texttt{toUpper}	& 2 & 3 & 5 & 1 & 0 & 0 & 601.04 & 133.97 & 735.13 \\ \hline
		\texttt{trim}		& 5 & 24 & 10 & 12 & 0 & 0 & 11.75 & 624.89 & 641.53 \\ \hline
		
		\multicolumn{10}{l|}{\cellcolor{black}\textcolor{white}{\textit{\textbf{$\mathcal{PS}$}}}}\\
		
		\texttt{charAt}		& 2 & 5 & 13 & 3 & 0 & 0 & 1.81 & 3.95 & 5.94\\ \hline
		\texttt{concat}		& 2 & 5 & 3 & 7 & 0 & 0 & 0.59 & 8.20 & 8.92\\ \hline
		\texttt{contains}	& 3 & 8 & 2 & 2 & 1 & 1 & 2.27 & 5.41 & 9.13\\ \hline
		\texttt{toLower}	& 2 & 5 & 5 & 3 & 0 & 0 & 2.75 & 4.03 & 6.91\\ \hline
		\texttt{toUpper}	& 2 & 5 & 5 & 3 & 0 & 0 & 1.73 & 3.99 & 5.85\\ \hline
		\texttt{trim}		& 2 & 5 & 11 & 3 & 0 & 0 & 1.74 & 6.58 & 8.52\\ \hline
		
		\multicolumn{10}{l|}{\cellcolor{black}\textcolor{white}{\textit{\textbf{$\mathcal{SH}$}}}}\\
		
		\texttt{charAt}		& 2 & 2 & 7 & 0 & 0 & 0 & 1.84 & 1.78 & 3.76\\ \hline
		\texttt{concat}		& 2 & 2 & 9 & 0 & 0 & 0 & 600.89 &8.07 & 609.30\\ \hline
		\texttt{contains}	& 3 & 8 & 7 & 3 & 1 & 1 & 2.56 & 6.15 & 10.39\\ \hline
		\texttt{toLower}	& 4 & 6 & 3 & 0 & 0 & 0 & 2.99 & 5.29 & 8.44\\ \hline
		\texttt{toUpper}	& 2 & 2 & 1 & 0 & 0 & 0 & 1.87 & 1.75 & 3.73\\ \hline
		\texttt{trim}		& 4 & 6 & 3 & 0 & 0 & 0 & 2.82 & 5.29 & 8.29\\ \hline
		
	\end{tabular}
\end{table*}
}
\forOOPSLA{
	\pankaj{
		Detailed log of experiments on synthesizing abstract transformers for abstract string domain  is available in the extended version \cite{amurthARXIV}.
	}
}

As it happens, the $\mathcal{SH}$ domain only supports a non-trivial abstract transformer for the
concrete function \texttt{concat};
for each of the other five functions, the best abstract transformer is
the trivial abstract transformer
$\lambda a.(\textrm{if }a=\bot \textrm{ then } \bot \textrm{ else } \top)$~\citep{TACAS:SAFEstr17,CC:Magnus14}.
\tool synthesized these transformers, too;
each took a nontrivial amount of time because \tool had to establish that the DSL could not express a more
precise abstract transformer.

In this section, we focus the discussion on the transformers that
produced interesting behaviors and challenges, and discuss the DSLs
that we used.

\subsubsubsection{Transformer for $\mathtt{contains}$ in  the $\charin$ Domain.}
The concrete function $\mathtt{contains(arg1,arg2)}$ returns \texttt{True}
if \texttt{arg2} is a
contiguous
substring of \texttt{arg1}, and \texttt{False} otherwise.
\tool takes 1,804.69s to synthesize the abstract transformer
$\abst{\texttt{contains}}{{\texttt{CI}}}$ shown in \figref{containsSafe}.
The transformer takes two abstract inputs in $\charin$, and returns an
abstract Boolean value in \texttt{AbsBool}.
\texttt{AbsBool} contains four elements, \texttt{BoolBot}, \texttt{BoolTrue}, \texttt{BoolFalse}, and \texttt{BoolTop},
and these elements  satisfy the partial order $\texttt{BoolBot} \sqsubseteq value \sqsubseteq \texttt{BoolTop}$,
where $value$ can be either $\texttt{BoolTrue}$ or $\texttt{BoolFalse}$.
\texttt{BoolTrue} and \texttt{BoolFalse} are incomparable.
\twrchanged{
The DSL we used to synthesize this transformer is
}
{
\begin{equation}
  \label{Eq:DSLForContainsCI}
  \begin{array}{@{\hspace{0ex}}r@{\hspace{1.0ex}}c@{\hspace{1.0ex}}l@{\hspace{0ex}}}
    \textit{Transformer} & ::= & \lambda \mathtt{a_1,a_2} . AB \\
                      AB & ::= & \texttt{ite}(B, AB, AB) \mid \texttt{BoolTop} \mid \texttt{BoolBot}  \mid \texttt{BoolTrue}\mid  \texttt{BoolFalse} \\
                       B & ::= & \texttt{isSubset}(LU, LU) \mid \texttt{size}(LU) \leq 1  \mid \texttt{isBot}(CI) \mid \texttt{isTop}(CI) \mid \texttt{isEmpty}(CI) \\
                         &     & \quad \mid \neg B \mid B\land B \mid B\lor B \\
                      CI & ::= & \mathtt{a_1} \mid \mathtt{a_2} \\
                      LU & ::= & CI.\texttt{l}\mid  CI.\texttt{u}
  \end{array}
\end{equation}
}
A program in this DSL computes an \texttt{AbsBool}, hence the initial nonterminal $AB$.
Other nonterminals denote the types Boolean ($B$), $\charin$ ($CI$), and $L$, $U$ values ($LU$).
This DSL contains a number of auxiliary functions, e.g., \texttt{isBot}, \texttt{isTop},  \texttt{isSubset}, which can be used to
inspect abstract values.
The auxiliary function \texttt{isBot} (resp.\ \texttt{isTop}) checks whether an abstract
value in $\charin$ is $\bot_\charin$ (resp.\ $\top_\charin$).
The function, \texttt{isSubset(a1,a2)} returns \texttt{True} iff \texttt{a1} is a subset of \texttt{a2}.
\texttt{isEmpty(a)} returns \texttt{True} iff \texttt{a} represents the empty set.

\begin{figure}
    \begin{subfigure}{.55\linewidth}
\begin{lstlisting}[language=C,numbers=left, escapechar=\$]
$\texttt{contains}^\sharp_{\texttt{CI}} (\mathtt{a_1}:CI) (\mathtt{a_2}:CI): \mathtt{AbsBool} =$
 ite(isBot($\mathtt{a_1.l,a_1.u}$)$\lor$isBot($\mathtt{a_2.l,a_2.u}$),
   boolBot,
$\textcolor{red}{\mathtt{[-]\ ite(isTop(a_1.l,a_1.u) \lor isTop(a_2.l,a_2.u),}}$ $\textcolor{green!50!black}{\mathtt{ //\ Bug}}$$\label{Li:topContainBug}$
$\textcolor{red}{\mathtt{[-]\ \ boolTop,}}$$\textcolor{green!50!black}{\mathtt{\quad\quad\quad\quad\quad\quad\quad\quad\quad\quad\quad\quad\quad\quad //\ Bug}}$$\label{Li:topContainBugTop}$
     ite($\neg$isSubset($\mathtt{a_2.l,a_1.u}$),
       boolFalse,$\label{line:retFalse}$
   $\textcolor{red}{\texttt{[-]}\ \mathtt{ite(size(a_2.u}) \leq 1 \land \mathtt{isSubset(a_2.u,a_1.l}),}$$\label{line:condTrue}$$\textcolor{green!50!black}{\texttt{ // Bug}}$$\label{Li:BuggyCondition}$
   $\textcolor{blue}{\mathtt{[+]\ ite(isEmpty(a_2),}}$ $\textcolor{green!50!black}{\qquad\qquad\qquad\qquad\qquad\ \hspace{0.2em}\texttt{// Fix}}$$\label{Li:FixIsEmpty}$
      $\qquad\mathtt{boolTrue},$$\label{Li:retTrue}$
     $\textcolor{red}{\mathtt{[-]\ \ boolTop))))}}$$\textcolor{green!50!black}{\mathtt{\quad\quad\quad\quad\quad\quad\quad\quad\quad\quad\quad\quad //\ Bug}}$$\label{Li:topContainBugTopTwo}$
     $\textcolor{blue}{\mathtt{[+]\ \ boolTop)))}}$$\textcolor{green!50!black}{\mathtt{\quad\quad\quad\quad\quad\quad\quad\quad\quad\quad\quad\quad \ //\ Fix}}$$\label{Li:topContainBugTopTwoFix}$
\end{lstlisting}
        \subcaption{Abstract transformers for \texttt{contains}.\label{Fi:containsSafe}}
    \end{subfigure}%
\hfill
    \begin{subfigure}{.45\linewidth}
\begin{lstlisting}[language=C,numbers=left]
$\texttt{trim}^\sharp_{\texttt{CI}} (\mathtt{a}:CI): CI =$
 ite(isBot(a.l,a.u),
   Bot,
   ite(isTop(a.l,a.u),
     Top,
     ite(size(a.u) $\leq$1 $\land$ containsSpace(a.u),
       $[\emptyset,\emptyset]$,
   $\texttt{\textcolor{red}{[-]\ a} \qquad\qquad\qquad\qquad\qquad\qquad \textcolor{green!50!black}{// Bug}}$
   $\texttt{\textcolor{blue}{[+]\ [removeSpace(a.l), a.u]} \quad \textcolor{green!50!black}{// Fix}}$
       )))
\end{lstlisting}
        \subcaption{Abstract transformers for \texttt{trim}.\label{Fi:trimSketch}}
    \end{subfigure}
    \caption{Bugs found and fixed in the $\charin$ domain for \texttt{contains} and \texttt{trim}.
    The lines in blue show how the synthesized transformers differ from the incorrect ones in {\safestr} (denoted by the lines in red).}
\end{figure}

\twrchanged{
When comparing the transformer $\texttt{contains}^\sharp_{\texttt{CI}}$ synthesized
using \tool (with \lineref{FixIsEmpty} in \figref{containsSafe}) to the one
implemented in {\safestr} (with \lineref{BuggyCondition} in \figref{containsSafe}),
we discovered that the latter was not sound.
}
The following example illustrates the problem.
Consider two abstract values
$\mathtt{a1 = [\{'a'\}, \{'a', 'b'\}]}$ and $\mathtt{a2 = [\{ \}, \{'a'\}]}$.
When the $\mathcal{CI}$ abstract transformer implemented in \safestr is applied to
$\mathtt{a1}$ and $\mathtt{a2}$, it returns $\mathtt{BoolTrue}$.
For $\mathtt{BoolTrue}$ to be the correct answer, every string in $\gamma(\mathtt{a2})$
must be a contiguous substring of every string in $\gamma(\mathtt{a1})$.
However, $\mathtt{``aaa"} \in \gamma(\mathtt{a2})$, and
$\mathtt{``ababa"} \in \gamma(\mathtt{a1})$, but
$\mathtt{``aaa"}$ is not a contiguous substring of $\gamma(\mathtt{a2})$.
Therefore, \safestr's $\mathcal{CI}$ transformer has a bug: it is unsound.
The transformer synthesized by \tool is sound (and a best $L$-transformer with respect
to the DSL given above).

\twrchanged{
Our inspection also revealed that $\texttt{contains}^\sharp_{\texttt{CI}}$ in {\safestr}
contained a precision bug:
$\mathtt{contains}_{CI}^\sharp$ should return \texttt{boolTrue} when
$a_1 = \top$ and $a_2$ is the empty string.
In \safestr, it returns \texttt{boolTop}, which is sound but imprecise.
In contrast, the transformer synthesized by \tool (without
\lineseqref{topContainBug}{topContainBugTop}, and
\lineref{topContainBugTopTwoFix} in place of \lineref{topContainBugTopTwo})
returns \texttt{boolTrue}:
$\mathtt{isSubset(a_2.l, a_1.u)}$ is \texttt{true}, and $\mathtt{a_2}$ is empty.
}


\Omit{
In the $\charin$ domain, it is not always possible to give a more precise answer than $\mathtt{BoolTop}$.
However, if $\mathtt{a_2.l}$ is \textit{not} a subset of $\mathtt{a_1.u}$, then we can surely say that
$\mathtt{a_1}$ does \textit{not} contain $\mathtt{a_2}$.
In this case, the synthesized transformer $\texttt{contains}^\sharp_{\texttt{CI}}$
correctly returns \texttt{BoolFalse},
while  the incorrect transformer implemented in {\safestr} returns \texttt{BoolTop},
when $\mathtt{a_1}$ and $\mathtt{a_2}$ are not $\mathtt{Bot}$.
}

\subsubsubsection{Transformer for ${\mathtt{trim}}$ in  the $\charin$ Domain.}
The function \texttt{trim} takes a string \texttt{s} and removes all the whitespace
at the beginning and the end of \texttt{s}.
\tool synthesizes the transformer $\texttt{trim}^\sharp_{\texttt{CI}}$ in \figref{trimSketch} in 641.53s.
The DSL used when synthesizing this transformer is
{
\begin{equation}
  \label{Eq:DSLForTrimCI}
  \begin{array}{@{\hspace{0ex}}r@{\hspace{1.0ex}}c@{\hspace{1.0ex}}l@{\hspace{0ex}}}
    \textit{Transformer} & ::= & \lambda \texttt{a} . CI \\
                      CI & ::= & \texttt{a}\mid [\emptyset, \emptyset] \mid [LU, LU]\mid \texttt{ite}(B, CI, CI) \\
                       B & ::= & \texttt{size}(LU) \leq 1 \mid  \texttt{isBot}(CI) \mid \texttt{isTop}(CI) \mid \texttt{containsSpace}(LU)|\ \neg B \mid B\land B \mid  B \lor B \\
                      LU & ::= & CI.\texttt{l}\mid  CI.\texttt{u}\mid  \texttt{removeSpace}(LU)
  \end{array}
\end{equation}
}
A program in the DSL returns an abstract string in the $\charin$ domain, hence the initial nonterminal is $CI$.
The DSL contains the following operators:
\texttt{size} returns the size of the argument set,
\texttt{containsSpace} returns \texttt{True} iff the argument set contains a whitespace character,
and \texttt{removeSpace} removes any whitespace character from the argument set.

When comparing the transformer synthesized using \tool to the one
implemented in {\safestr} (blue and red lines in \figref{trimSketch}), we discovered that the latter was not sound.
Consider the abstract input value \texttt{absArg} = \texttt{[\{\sq \textvisiblespace \sq,\sq a\sq\}, \{\sq \textvisiblespace \sq,\sq a\sq,\sq b\sq,\sq c\sq\}]} for which the concretization contains---among other values---the concrete string \texttt{s = "\textvisiblespace \textvisiblespace abc\textvisiblespace \textvisiblespace"}.
On this input, $\texttt{trim}^\sharp_{\mathtt{CI}}$ returns as output the
abstract value \texttt{absArg'}, which is same as \texttt{absArg}.
However, \texttt{trim(s)="abc"}, whereas $\texttt{"abc"} \notin \gamma(\texttt{absArg'})$
(because $\texttt{\sq \textvisiblespace \sq}$ is an element of the must-set of
\texttt{absArg'}).
Consequently, the abstract-transformer implementation in {\safestr} is unsound.
The transformer synthesized by {\tool} ($\texttt{trim}^\sharp_{\mathtt{synCI}}$) does not have this issue.

\subsubsubsection{Transformer for $\mathtt{trim}$ in the $\mathcal{PS}$ Domain.}

\begin{figure}
\begin{minipage}{.45\columnwidth}
	\begin{tabular}{c}
\begin{lstlisting}[language=Scala,numbers=left]
$\texttt{trim}^\sharp_{\mathtt{PS}} (\mathtt{a}:PS): PS =$
 ite(isBot(a.p,a.s),
   BOT,
   ite(isTop(a.p,a.s),
    TOP,
$\texttt{\textcolor{red}{[-] [trimStart(a.p), trimEnd(a.s)]} \textcolor{green!50!black}{// Bug}}\label{Li:PSBug}$
$\texttt{\textcolor{blue}{[+] [trim(a.p), trim(a.s)]} \textcolor{green!50!black}{\qquad\qquad// Fix}}\label{Li:PSFix}$
))
\end{lstlisting}
    \end{tabular}
    \captionof{figure}{Abstract transformers for \texttt{trim} in the $\mathcal{PS}$ domain.\label{Fi:trimPSsketch} \label{Fi:trimPSsafe}}
\end{minipage}
\hspace{.5cm}
\begin{minipage}{.48\columnwidth}
	\begin{tabular}{c}
\begin{lstlisting}[language=Scala,numbers=left]
concat$^\sharp \mathtt{(a:Long) (b:Long): Long} =$
  r $\gets$ reverse(b); c $\gets$ 0; i $\gets$ 0
  WHILE i < b
    r $\gets$rotateLeft(r, 1)
    IF (a $\mathbin{\&}$ r) $\neq 0$ THEN
$\mathtt{\textcolor{red}{[-]\quad\ c \gets c  \mathbin{|} (1 << i)}\textcolor{green!50!black}{\quad  // \safestr}}\label{Li:hashSafeOp}$
$\mathtt{\textcolor{blue}{[+]\quad\ c \gets c \mathbin{\string^} (1 << i)}\textcolor{green!50!black}{\quad //\normalfont\tool}}\label{Li:hashAmurthOp}$
    i $\gets$ i + 1
  RETURN c
\end{lstlisting}
	\end{tabular}
	\captionof{figure}{Abstract transformers for \texttt{concat} in the $\mathcal{SH}$ domain.\label{Fi:hashConcat}}
\end{minipage}
\end{figure}

An element of the $\mathcal{PS}$ domain is a pair $[p,s]$ describing the longest common prefix $p$ and suffix $s$ of a set of strings.
Consider the abstract value $a=[\texttt{"\textvisiblespace b\textvisiblespace"}, \texttt{"\textvisiblespace b\textvisiblespace"}]$, and the string $\texttt{"\textvisiblespace b\textvisiblespace"}\in \gamma(a)$.
A most precise transformer for \texttt{trim} on input $a$ should output $a'=[\texttt{"b"}, \texttt{"b"}]$.

The transformer $\mathtt{trim}^\sharp_{\mathtt{PS}}$ implemented in {\safestr} is unsound.
It incorrectly produces the output $a'=[\texttt{"b\textvisiblespace"}, \texttt{"\textvisiblespace b"}]$,
whose concretization fails to contain the concrete value
$\texttt{trim}(\texttt{"\textvisiblespace b\textvisiblespace"})=\texttt{b}$.
This bug is due to the statement at \lineref{PSBug} of~\figref{trimPSsketch}.
Using the DSL defined in \eqref{TrimPSDSL}, \tool is able to synthesize (in 8.52s)
a correct version of $\mathtt{trim}^\sharp_{\mathtt{PS}}$:
in \figref{trimPSsketch}, \lineref{PSBug} is replaced by \lineref{PSFix}.
\begin{equation}
  \label{Eq:TrimPSDSL}
  \begin{array}{@{\hspace{0ex}}r@{\hspace{1.0ex}}c@{\hspace{1.0ex}}l@{\hspace{0ex}}}
	\textit{Transformer} & ::= & \lambda \texttt{a} . PS \\
	PS & ::= & \texttt{a}\mid [\texttt{\sq \sq, \sq \sq}] \mid [LU, LU]\mid \texttt{ite(B, PS, PS)}   \mid BOT \mid TOP \\
	 B & ::= &   \texttt{isBot(PS)} \mid \texttt{isTop(PS)} \\
	LU & ::= & PS.\texttt{p}\mid  PS.\texttt{s} \mid \texttt{trim(LU)} \mid \texttt{trimStart(LU)} \mid \texttt{trimEnd(LU)}
  \end{array}
\end{equation}


\subsubsubsection{Transformer for $\mathtt{concat}$ in  the $\mathcal{SH}$ Domain.}

\begin{wrapfigure}{R}{.55\linewidth}
\begin{lstlisting}[numbers=left]
#define W 64

bit[W] absConcat(bit[W] a, bit[W] b){
  bit[W] r = {| a | b | reverse(a) | reverse(b)|};
  bit[W] one = {1};
  bit[W] c = {0};
  bit[W] cond;
  for(int i = 0; i < W; i++){
    one = {1};
    r = {| rotateLeft(r, 1) | rotateRight(r, 1) |}; $\label{Li:HashDSLChoice}$
    cond = {| (a & r) | (a | r) | (a ^ r) |};
    if( (cond) != (bit[W]){0}){
      c = {| (c & rotateLeft(one, i))
           | (c | rotateLeft(one, i))
	   | (c ^ rotateLeft(one, i))
	   | (c & rotateRight(one, i))
	   | (c & rotateRight(one, i))
	   | (c ^ rotateRight(one, i)) |};
    }
  }
  return c;
}
\end{lstlisting}
	\caption{Sketch to synthesize abstract transformer for $\mathtt{concat}^\sharp$ in $\mathcal{SH}$\label{Fi:concatSk}}
\end{wrapfigure}

The concrete function \texttt{concat} takes two concrete strings and returns their concatenation.
\twrchanged{
Pseudocode for the $\mathtt{concat^\sharp}$ transformer used in \safestr
is shown in \figref{hashConcat} with \lineref{hashSafeOp} (and without \lineref{hashAmurthOp}).
}
{\safestr} uses a universe $U$ of values of size $64$ (the range of the hash function).
The abstract transformer uses a 64-bit \texttt{long} value $a$ as a bit-vector encoding of
a subset $A\subseteq U$---i.e., the $i$-th bit of $a$ is $1$ iff $i \in A$.
\twrchanged{
\tool synthesizes the abstract transformer $\mathtt{concat^\sharp_{syn}}$ whose
pseudocode is shown in \figref{hashConcat} with \lineref{hashAmurthOp} (and without \lineref{hashSafeOp}).
}
The  DSL used to synthesize this transformer
uses a different structure than the ones described above.
\twrchanged{
In particular, it is a sketch of the loop that we expect the target
function to contain, as shown in \figref{concatSk}.
From the options specified within \texttt{\{|...|...|...|\}},
the Sketch synthesizer selects an option that makes the resulting program
consistent with the specification.
For example, at \lineref{HashDSLChoice}, Sketch selects from among
the provided options \texttt{rotateLeft(r,1)} and \texttt{rotateRight(r,1)}.
}
The sketch can then be completed using bitwise operators---i.e., not
(\texttt{$\neg$}), or (\texttt{|}), and (\texttt{\&}), xor (\texttt{\^}),
left shift (\verb|<<|), right shift (\verb|>>|), left-rotate, right-rotate, and \texttt{reverse}.
\tool takes 609.30s to synthesize the abstract transformer for
\texttt{concat} in the $\mathcal{SH}$ domain (see
\tableref{strdom-synthesis}).

This transformer is particularly interesting because it contains
complex logic, and an implementation trick that is hard to reason
about.
While we had to provide the overall structure of the program, \tool
could fill in the tricky implementation details automatically.
This type of approach has been used before in program synthesis.
For example, the original motivation for Sketch itself
was to synthesize tricky implementation details
of user-provided implementation sketches for bit-vector operations.
This particular example shows that \tool can synthesize highly
non-trivial abstract transformers.

\subsubsection{Performance and Precision of the Synthesized Transformers in a Program Analyzer}
We compared the performance and precision of the hand-written transformers in \safestr with
the transformers synthesized by {\tool}.
The three bugs found in \safestr were fixed for this experiment.
We ran all the benchmark verification problems provided in \safestr,
and collected the same precision metric (``\textit{imprecision index}'' \citep{TACAS:SAFEstr17})
used to evaluate \safestr.
For each benchmark, we compared the performance with respect to the time,
the number of fixpoint iterations of the analysis, the number of reachable program states,
and the precision metric from \safestr.

\begin{figure}
	\centering
	\includegraphics[scale=0.35]{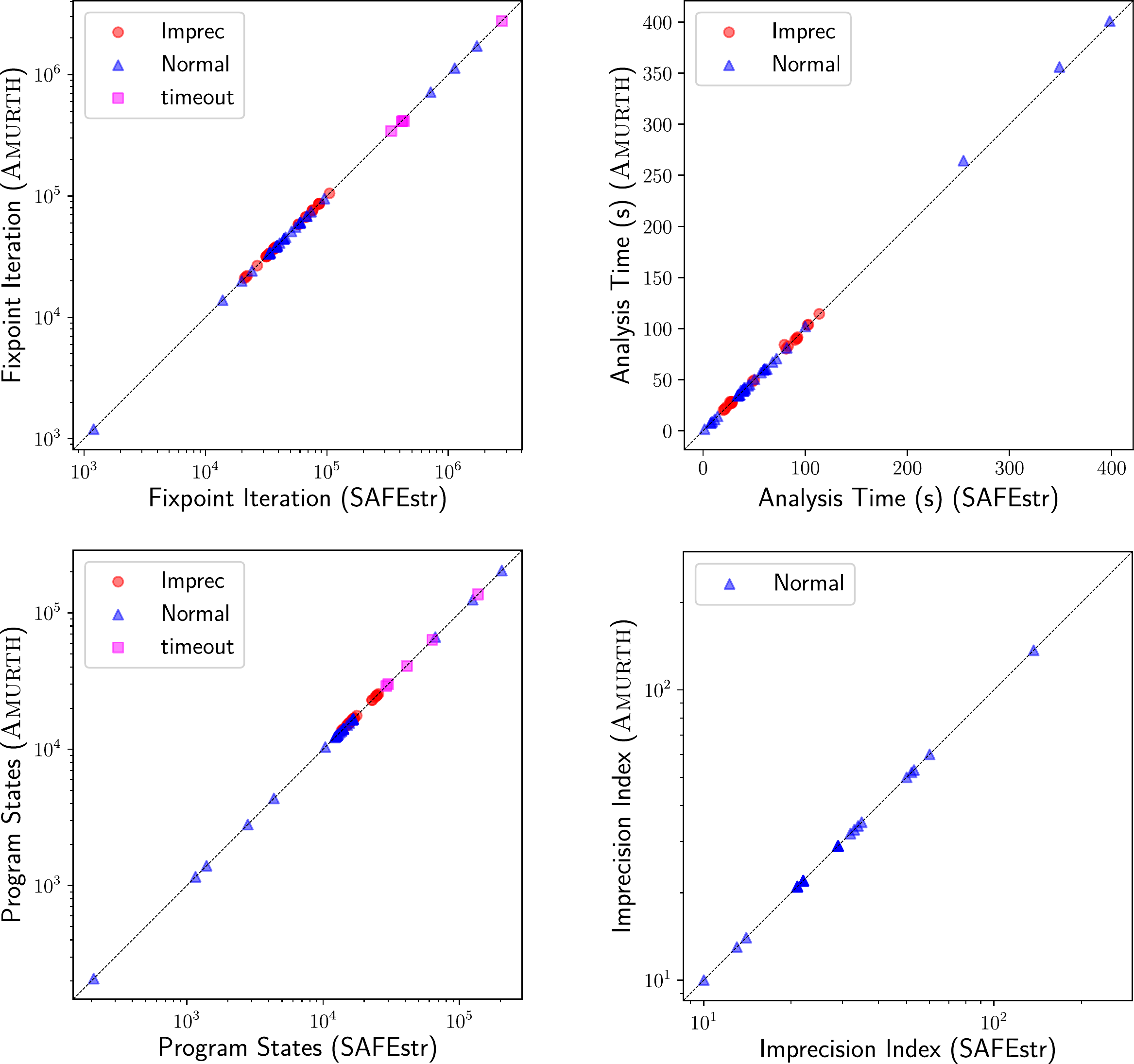}
	\caption{Performance and precision of the synthesized transformers with the product domain $\csdom \times \charin$.
                 \twrchanged{In the left-hand plots, all examples labeled ``timeout'' exceeded the timeout threshold
                 with both the hand-written \safestr transformers and the transformers synthesized by \tool.
                 (There were no examples in which one set of transformers exceeded the timeout threshold and the other
                 set did not.)
}
		\label{Fi:CsCiEvaluation}}
\end{figure}

The scatter plots in
\forARXIV{
	Figs.~\ref{Fi:CsCiEvaluation}, \ref{Fi:COvaluation}, \ref{Fi:SSKEvaluation} and \ref{Fi:CoPsEvaluation}
}
\forOOPSLA{
	Figs.~\ref{Fi:CsCiEvaluation} (plots for other domains are available in extended version~\cite{amurthARXIV})
}
show the data from runs using the {\safestr}
transformers on the $x$-axis, and the data from runs using the
transformers synthesized by \tool on the $y$-axis.
\twrchanged{
For both the hand-written transformers in \safestr and the
transformers synthesized by \tool, each run of the analyzer is given a
timeout threshold of 600 seconds.
}
An analysis run can also terminate with a \safestr imprecision-trigger
exception (``Imprec'') if the imprecision becomes too great. The
following symbols are used in the plots to show the status of an
analysis:
a magenta square (\nsquare{magenta}) shows that the analysis timed out;
a blue triangle (\ntriangle{blue}) indicates normal termination;
and a red circle (\ncircle{red}) indicates termination due to an imprecision-trigger exception.
The plots show that the transformers synthesized by \tool have the same performance
and precision as the ones implemented in \safestr (i.e., all points lie essentially on the diagonal line).
\twrchanged{
There were no examples of a run using the hand-written \safestr
transformers that timed out, but the corresponding run with the
transformers synthesized by \tool completing within the time limit, or
vice versa.
}

For the $\mathcal{SH}$ domain, the abstract transformer synthesized by  \tool for \texttt{concat} (\lineref{hashAmurthOp} of \figref{hashConcat}) is equivalent to the hand-written transformer in \safestr (\lineref{hashSafeOp} of \figref{hashConcat}). Hence, we do not show plots for the $\mathcal{SH}$ domain.

Because a singleton set of strings cannot be represented precisely in
the $\mathcal{CI}$ and $\mathcal{PS}$ domains, to provide a better
basis for comparing precision, we used instead their direct products
with $\mathcal{CS}$ (i.e., $\csdom \times \charin$ and $\csdom \times
\psdom$).
See \figref{CsCiEvaluation} 
\forARXIV{
and \ref{Fi:CoPsEvaluation}.
}
\forOOPSLA{
	????
}

\forARXIV{
\begin{figure}
	\centering
	\includegraphics[scale=0.35]{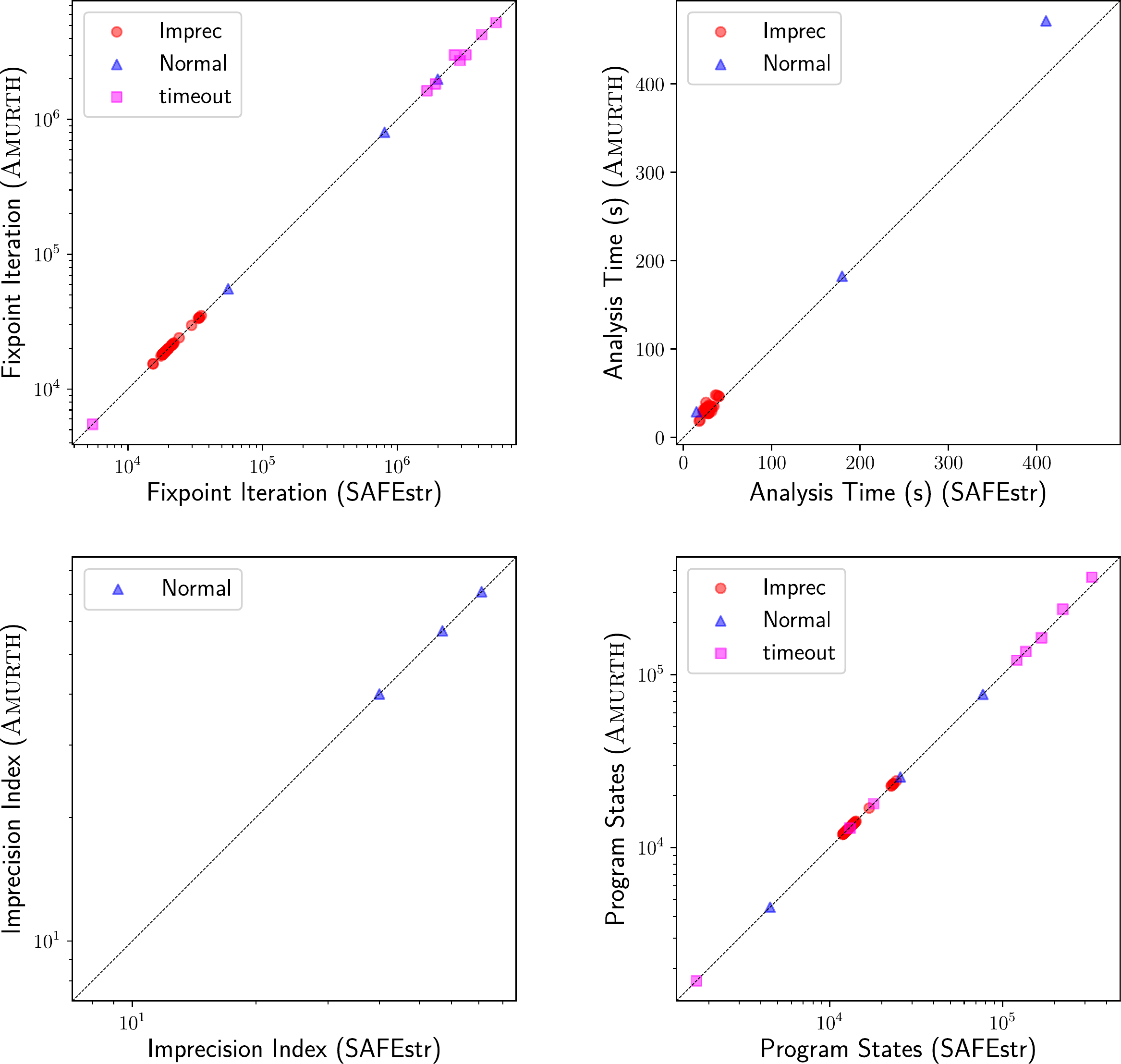}
	\caption{Performance and precision of the synthesized transformers with the $\csdom$ domain.
		(\nsquare{magenta}~: timeout; \ntriangle{blue}~: normal termination; \ncircle{red}~: termination due to imprecision trigger)
		\label{Fi:COvaluation}}
\end{figure}

\begin{figure}
	\centering
	\includegraphics[scale=0.35]{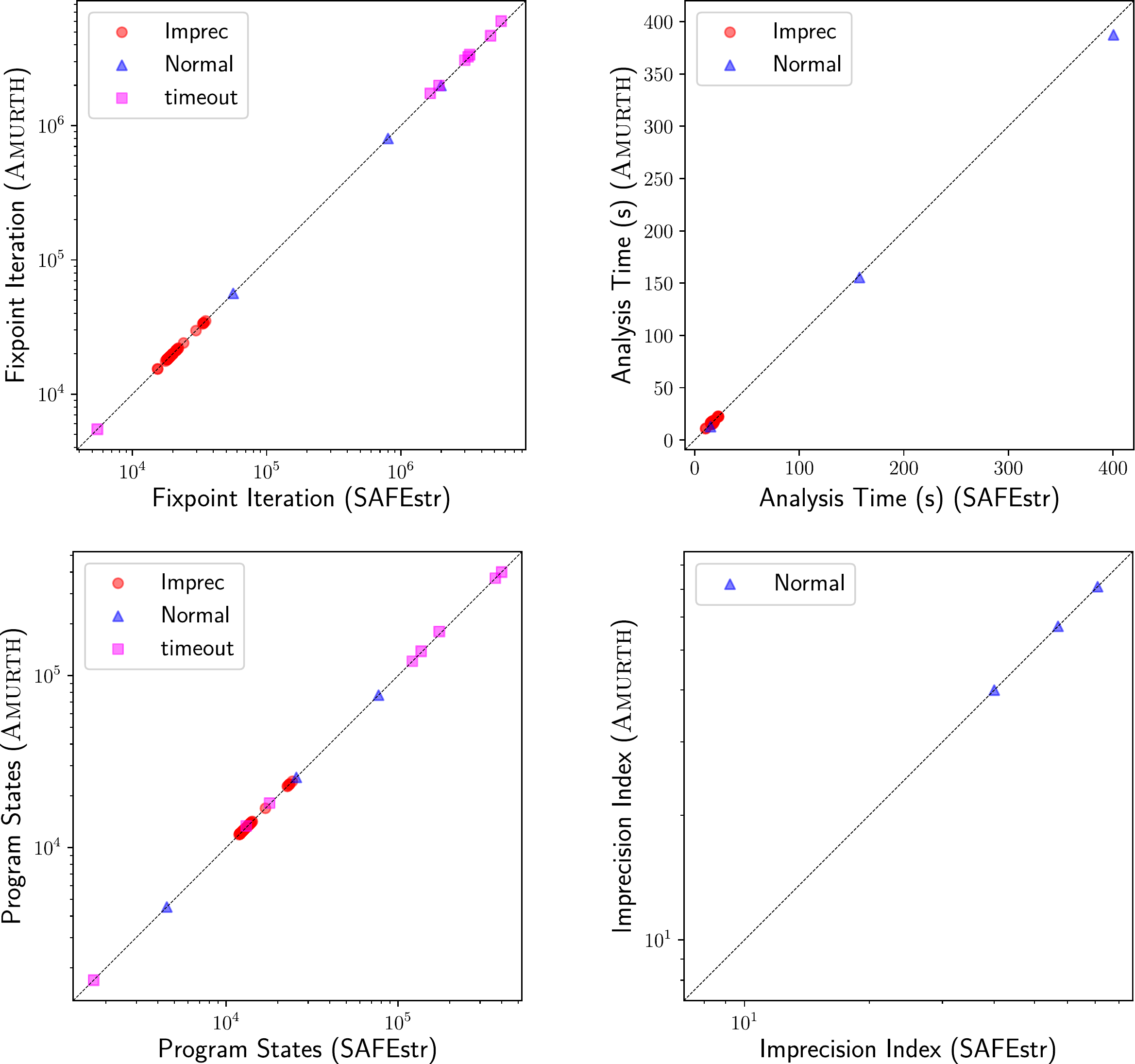}
	\caption{Performance and precision of the synthesized transformers  with the $\ssk$ domain $(k = 3)$.
		\label{Fi:SSKEvaluation}}
\end{figure}

\begin{figure}
	\centering
	\includegraphics[scale=0.35]{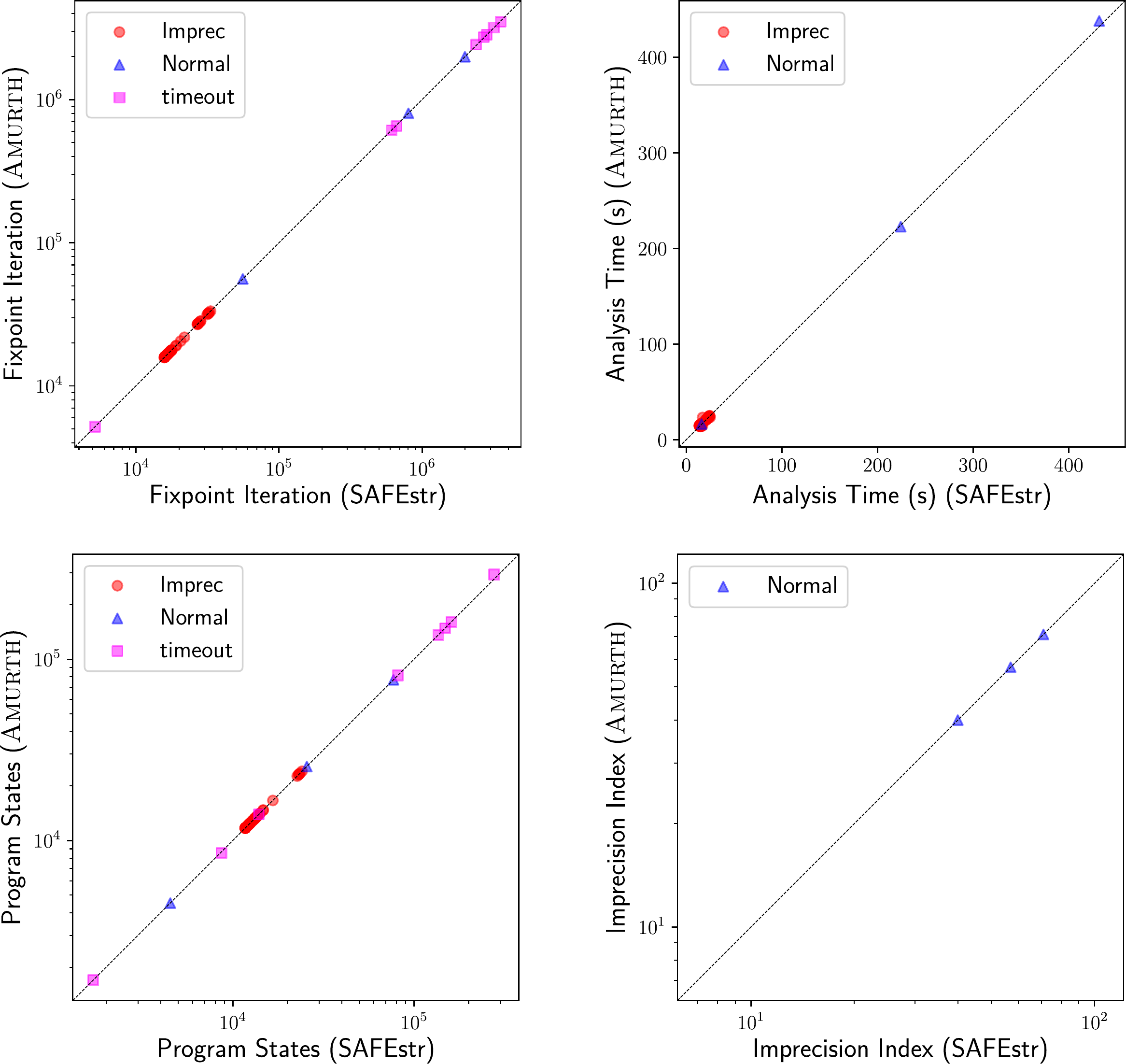}
	\caption{Performance and precision of the synthesized transformers with the product domain $\csdom \times \psdom$.
		\label{Fi:CoPsEvaluation}}
\end{figure}
}

\smallskip
\noindent
\begin{mdframed}[innerleftmargin = 3pt, innerrightmargin = 3pt, skipbelow=-0.25em]
  \textbf{Finding [RQ1]:}
  The time taken by \tool to synthesize an $L$-transformer across all
  of the string-transformer experiments varies between 3.73s and 1,983.83s.
	\FuzzOut{
  \newline
  \textbf{Finding [RQ2]:}
	For each function, running tool with a logical specification and a closed-box specification of the function resulted in equivalent
				transformers.}
  \newline
  \textbf{Finding [RQ2]:}
  A manual comparison of the automatically-generated and
  manually-written transformers revealed that three of the manually
  written transformers in \safestr were unsound:
  $\mathtt{contains}^\sharp$ for the $\charin$ domain, and
  $\mathtt{trim}^\sharp$ for $\charin$ and $\mathcal{PS}$.

  Figs.~\ref{Fi:CsCiEvaluation} 
  \forARXIV{
\ref{Fi:COvaluation}, \ref{Fi:SSKEvaluation} and \ref{Fi:CoPsEvaluation} 
}
\forOOPSLA{
???
}
show that the abstract transformers synthesized by \tool
  for the six string operations are
  empirically indistinguishable---in terms of analysis time and precision---from
  the manually written ones used in \safestr
  (after the three buggy \safestr transformers were fixed).
\end{mdframed}

\subsection{Case Study 2: Transformers for Three Fixed-Bitwidth Interval Domains}
\label{Se:casestudy1}

In this study, we used {\tool} to synthesize abstract transformers in
the three fixed-bitwidth interval domains described in
\sectref{IntervalDomains}, for nine different mathematical and logical
operations.
Concrete arithmetic operations are performed in modular arithmetic
(sometimes known as ``machine-integer arithmetic'').
Each domain represents a set of fixed-bitwidth integers, and is
parameterized on $w$, which denotes the number of bits in a
represented integer.

\subsubsection{Fixed-Bitwidth Interval Domains}
\label{Se:IntervalDomains}

Our study considered the three fixed-bitwidth interval domains summarized below \citep{APLAS:WrInterval12}.

\forOOPSLA{
\begin{wrapfigure}{r}{.45\linewidth}
	\vspace{-5mm}
}
\forARXIV{
\begin{figure}
}
	\includegraphics[scale=0.45]{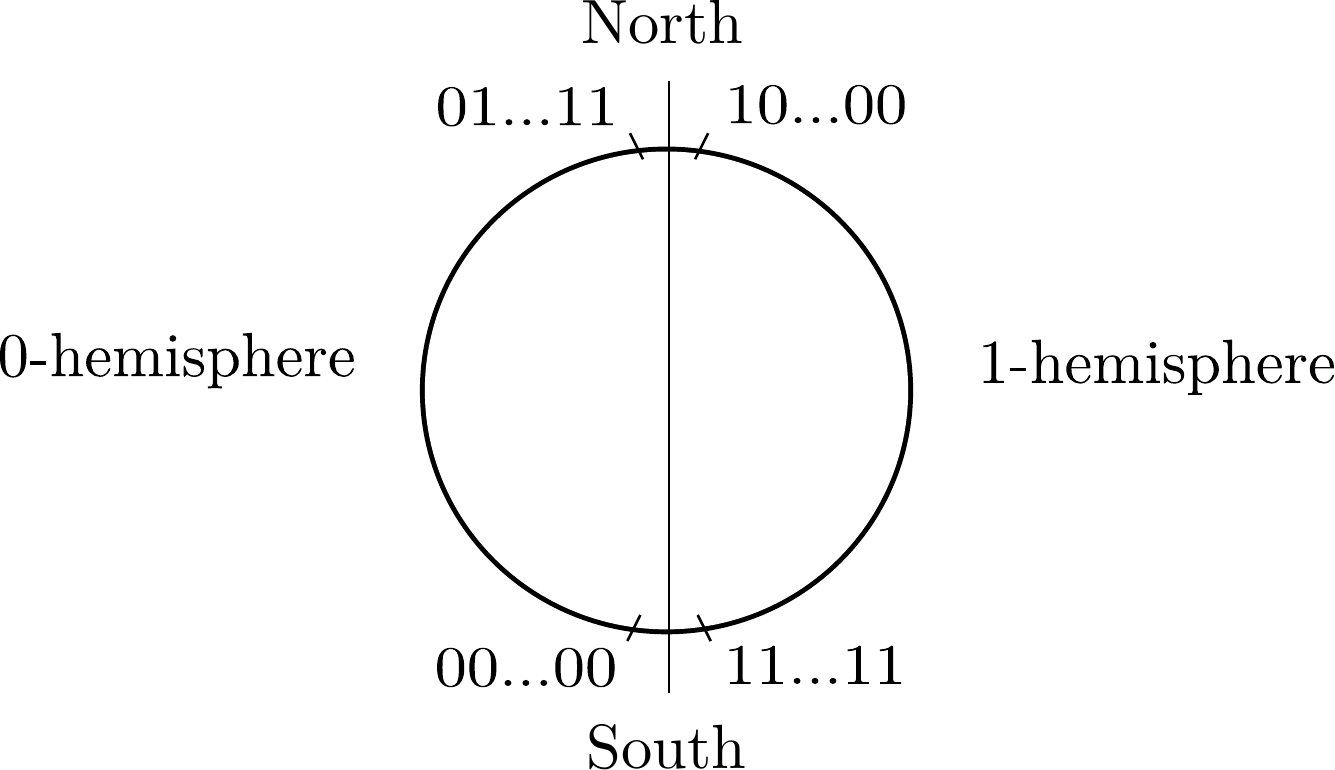}
	\caption{Wrapped Interval ($\wintv$) number circle.\label{Fi:wrappedIntervalNumCircle}}
\forARXIV{
	\end{figure}
}
\forOOPSLA{
\end{wrapfigure}
}

\subsubsubsection{Unsigned-Integer Intervals ($\uintv$). }
An element in the $w$-bit unsigned-integer interval domain is either $\bot$, which denotes the empty set,
or is from the set $\{[a,b] \mid 0 \leq a \leq b < m\}$,
where $m = 2^w$.
The concretization function ($\gamma_{\mathtt{u}}$) is defined as, $\gamma_{\mathtt{u}}([a, b]) = \{a, a+1,\dots,b-1,b\}$.

\subsubsubsection{Signed-Integer Intervals ($\sintv$). }
An element in the $w$-bit signed-integer interval domain is either $\bot$, which denotes the empty set,
or is from the set $\{[a,b] \mid -m \leq a \leq b < m \}$, where $m = 2^{w-1}$.
Negative numbers are interpreted in their two's-complement representation. The concretization  function $\gamma_s$ is defined as, $\gamma_s([a,b]) = \{a, a+1,\dots,b-1, b\}$.

\subsubsubsection{Wrapped Intervals ($\wintv$). }
An element in the wrapped interval is either $\bot$, which denotes the empty set,
$\top$, which represents the set $\{ -2^{w-1} \leq a < 2^{w-1} \}$,
or it is represented by $[a,b]$, where $a, b$ are $w$-bit bit-vectors such that
$a \not\equiv (b+1)\ \mathtt{mod}\ 2^w$.
This domain is a sign-agnostic domain.
Wrapped intervals are permitted to cross either or both of
the ``North pole'' and ``South pole'' shown in
\figref{wrappedIntervalNumCircle}.\footnote{
	In contrast, unsigned-integer and signed-integer intervals are allowed to cross only
	the North and South poles, respectively.
}
The concretization function ($\gamma_w$) for the wrapped interval domain is
defined as follows, where $\leq_l$ is lexicographic ordering on bit-vectors.
\begin{equation}
\label{Eq:gammaWrapped}
\gamma_w([a,b]) = \begin{cases}
\{a,\dots,b\}                        & \text{if } a \leq_l b  \\
\{0^w,\dots,b\} \cup \{a,\dots,1^w\} & \text{otherwise}.
\end{cases}
\end{equation}

For example, for $3$-bit intervals,
$\gamma([111, 101]) = \{000,001,010,011,100,101,111\}$ illustrates the second case of the \eqref{gammaWrapped}.

\subsubsection{Abstract Transformers for Fixed-Bitwidth Interval Domains}

\begin{table}
	\caption{Time, in seconds, to synthesize abstract transformers for the fixed-bitwidth interval domains.
		Templates were used in synthesizing the abstract transformers of the shaded cells
		in the table.\label{Ta:intervals_synthesis}
	}
	\centering
	\footnotesize
	
	\begin{tabular}{l|Hc rr|rrrrrrr}
		\textbf{Domain}& \textbf{mode} & \multicolumn{3}{c|}{\textbf{Arith.\ Ops.}}
		& \multicolumn{6}{c}{\textbf{Bitwise Ops.}}\\ \hline\hline
		& & \multicolumn{1}{c}{add} & \multicolumn{1}{c}{sub} & \multicolumn{1}{c|}{mul} &
		\multicolumn{1}{c}{and} & \multicolumn{1}{c}{or} & \multicolumn{1}{c}{xor} &
		\multicolumn{1}{c}{shift left} & \makecell[c]{arithmetic\\ shift right}
		& \makecell[c]{logical\\ shift right}\\ \hline

		{\textit{Unsigned}} 
		
		& $\Phi$   &  133.70 & 99.07 & 1,449.85 & 14.54 & 17.36 & 1,095.94  & 12.52 & 6.15 & 12.88 \\
		

		\hline
		
		{\textit{Signed}} 
		
		& $\Phi$ &607.63 & 213.17 & 1,287.83 &\cellcolor{gray!25!white} 284.42 & \cellcolor{gray!25!white}958.16 &\cellcolor{gray!25!white} 1,234.94 & 11.78 & 7.61 & 4.23 \\
		

		\hline
		
		{\textit{Wrapped}} 
		
		& $\Phi$ & 858.60 & 739.47 &\cellcolor{gray!25!white} 880.12 &\cellcolor{gray!25!white} 1,360.98&\cellcolor{gray!25!white} 1,311.37 &\cellcolor{gray!25!white} 962.13 & 198.90 & 598.62 & 548.72\\
		

		\hline
		
	\end{tabular}
	
\end{table}

For each of the three domains, we used {\tool} to synthesize abstract
transformers for nine operations (\texttt{add}, \texttt{sub},
\texttt{mul}, \texttt{and}, \texttt{or}, \texttt{xor}, \texttt{shl},
\texttt{ashr}, and \texttt{lshr}).
{\tool} takes 4.23s to 1,449.85s to synthesize
a best $L$-transformer;
see \tableref{intervals_synthesis}.
\forARXIV{
\tableref{IntvDomLogicSpecLog} shows the detailed experimental results for
abstract-transformer synthesis in the fixed-bitwidth interval domains.

\begin{table*}[htbp]
	\caption{Table showing detailed results for the fixed-bitwidth interval-domain experiments. \textbf{\#MaxSat} and \textbf{\#Dropped} are 0 for all entries.\label{Ta:IntvDomLogicSpecLog}}
	\begin{tabular}{|r|r|r|r|r|r|r|r|}
		\hline
		$f$ &\#SO&\#PO&\#PosEx&\#NegEx&SO Time (s)&PO Time (s)&Time (s)\\ \hline \hline
		\multicolumn{8}{l|}{\cellcolor{black}\textcolor{white}{\textit{\textbf{Unsigned}}}}\\
		\hline
		\texttt{add}	&2&7&11&25&1.79&130.28&133.70 \\
		\hline
		\texttt{sub}	&2&6&11&24&1.72&97.08&99.07 \\
		\hline
		\texttt{mul}		&2&35&3&33&10.78&1,434.62&1,449.85 \\
		\hline
		\texttt{and}		&2&5&7&3&1.80&12.56&14.54 \\
		\hline
		\texttt{or}		&2&4&0&2&1.82&15.39&17.36 \\
		\hline
		\texttt{xor}		&2&15&6&13&4.51&1,090.28&1,095.94 \\
		\hline
		\texttt{shl}	&2&5&7&5&1.81&10.51&12.52 \\
		\hline
		\texttt{ashr}	&2&3&1&7&1.82&4.16&6.15 \\
		\hline
		\texttt{lshr}	&2&5&3&7&2.89&9.79&12.88 \\
		\hline
		\multicolumn{8}{l|}{\cellcolor{black}\textcolor{white}{\textit{\textbf{Signed}}}}\\
		\hline
		\texttt{add}	&2&3&11&21&4.62&602.87&607.63 \\
		\hline
		\texttt{sub}	&2&8&11&26&11.03&200.60&213.17 \\
		\hline
		\texttt{mul}		&3&33&12&29&255.95&1,028.39&1,287.83 \\
		\hline
		\texttt{and}		&2& 3 & 17 & 1 & 130.15 & 153.98 & 284.42 \\
		\hline
		\texttt{or}		& 3 & 45 & 18 & 41 & 44.35 & 910.63 & 958.16 \\
		\hline
		\texttt{xor}		& 2 & 3 & 17 & 1 & 602.25 & 632.16 & 1,234.94 \\
		\hline
		\texttt{shl}	&2&4&3&4&1.90&9.43&11.78 \\
		\hline
		\texttt{ashr}	&2&4&5&6&1.94&5.49&7.61 \\
		\hline
		\texttt{lshr}	&2&2&3&4&1.84&2.07&4.23 \\
		\hline
		\multicolumn{8}{l|}{\cellcolor{black}\textcolor{white}{\textit{\textbf{Wrapped}}}}\\
		\hline
		\texttt{add}	&2&12&13&14&61.24&796.45&858.60 \\
		\hline
		\texttt{sub}	&2&12&13&14&14.46&724.19&739.47 \\
		\hline
		\texttt{mul}		&4 &12 &10 &8  &6.22 & 870.79& 880.12 \\
		\hline
		\texttt{and}		&2&5&5&3&623.38&737.12&1,360.98 \\
		\hline
		\texttt{or}		&2&5&5&3&621.31&689.64&1,311.37 \\
		\hline
		\texttt{xor}		&2&8&5&8&90.81&870.45&962.13 \\
		\hline
		\texttt{shl}	&2&7&19&5&11.71&186.53&198.90 \\
		\hline
		\texttt{ashr}	&2&8&7&10&450.54&147.31&598.62 \\
		\hline
		\texttt{lshr}	&4&19&9&19&276.20&270.80&548.72 \\
		\hline
	\end{tabular}
\end{table*}
}
\forOOPSLA{
	\pankaj{
	Detailed experimental results for abstract-transformer synthesis in the fixed-bitwidth interval domains is available in the extended version of this work~\cite{amurthARXIV}.
	}
}
While in most cases, merely providing a DSL grammar was enough (indicated by the unshaded cells in \tableref{intervals_synthesis}), for some of the more involved transformers, we had to provide a \emph{template}---a sketch of the high-level implementation---and \tool was able to fill in the details (see the shaded cells).
The transformers for which we needed an implementation sketch are
discussed in 
\twrchanged{
	\sectref{absTransformXor}.
} 

When we tried to synthesize the abstract transformers for
$\texttt{xor}$ for the $\sintv$ and $\wintv$ domains using the auxiliary functions used by
\citeauthor{APLAS:WrInterval12}, \tool failed.
A closer examination revealed that there was a bug in the \citeauthor{APLAS:WrInterval12} implementation
of the \texttt{minAnd} auxiliary function: in two places, they used a bitwise negation($\sim$) instead
of an arithmetic negation ($-$) as shown in \figref{buggyMinAnd}.
After we fixed this bug, all of the $L$-transformers synthesized by \tool
are semantically equivalent to those provided in the implementation that
accompanies the \citeauthor{APLAS:WrInterval12} paper [\citeyear{APLAS:WrInterval12}].

\begin{figure}
	\begin{tabular}{c}
	\begin{lstlisting}
llvm::APInt minAnd(llvm::APInt a, const llvm::APInt& b, llvm::APInt c, const llvm::APInt& d) {
    llvm::APInt m = llvm::APInt::getOneBitSet(a.getBitWidth(), a.getBitWidth() - 1);
    while (m != 0) {
        if ((~a & ~c & m)) {
        $\textcolor{red}{\texttt{[-]\ llvm::APInt temp = (a | m)}\ \texttt{\&}\ \mbox{\textasciitilde}m}$; $\textcolor{green!50!black}{\texttt{// Bug }}$
        $\textcolor{blue}{\texttt{[+]\ llvm::APInt temp = (a | m)}\ \texttt{\&}\ -m}$; $\textcolor{green!50!black}{\texttt{// Fix}}$
            if (temp <= b) {
                a = temp;
                break;}
        $\textcolor{red}{\texttt{[-]\ temp = (c | m)}\ \texttt{\&}\ \mbox{\textasciitilde}m}$; $\textcolor{green!50!black}{\texttt{// Bug}}$
        $\textcolor{blue}{\texttt{[+]\ temp = (c | m)}\ \texttt{\&}\ -m}$; $\textcolor{green!50!black}{\texttt{// Fix}}$
            if (temp <= d) {
                c = temp;
                break;}
        }
        m = m.lshr(1);
    }
    return a & c;
}
	\end{lstlisting}
	\end{tabular}
	\caption{Buggy implementation of \texttt{minAnd} from the \citeauthor{APLAS:WrInterval12} implementation.\label{Fi:buggyMinAnd}}
\end{figure}

\subsubsection{Abstract transformer for multiplication for $\uintv$ and $\sintv$.}
\label{Se:AbstractTransformerForMultiplication}

For this operation, whenever there is a possibility of an overflow, the transformer returns $\top$.
An overflow is detected as follows:

\begin{equation*}
\begin{array}{@{\hspace{0ex}}r@{\hspace{1.00ex}}c@{\hspace{1.0ex}}c@{\hspace{1.0ex}}l@{\hspace{0ex}}}
  \abst{\texttt{overflow\_mul}}{}(a_1: \uintv, a_2: \uintv)
       & = &      & \texttt{overflows}(a_1.l * a_2.l) \lor \texttt{overflows}(a_1.l * a_2.r) \\
       &   & \lor & \texttt{overflows}(a_1.r * a_2.l) \lor \texttt{overflows}(a_1.r * a_2.r)
\end{array}
\end{equation*}

\noindent
where $\texttt{overflows}(x * y)$ returns \texttt{true} whenever the unsigned multiplication of $x$ and $y$ overflows. A similar case holds for the case of signed multiplication, but in this case, the overflow is checked with an overloaded version of \texttt{overflows()} that checks overflows of signed multiplications.

To complete the transformer, we used \tool to synthesize the case where there was no overflow.
In essence, we used the following transformer template:
\begin{equation}
\abst{\texttt{mul}}{}(a_1, a_2) =
\begin{cases}
\top & \abst{\texttt{overflow\_mul}}{}(a_1, a_2) = true \\
\abst{\texttt{mul}}{no} & \abst{\texttt{overflow\_mul}}{}(a_1, a_2) = false
\end{cases}
\end{equation}

We used the following DSL for both the signed and unsigned domains:
\begin{equation}
\label{Eq:LForIntervalDomainMul}
\begin{array}{@{\hspace{0ex}}r@{\hspace{1.0ex}}c@{\hspace{1.0ex}}l@{\hspace{0ex}}}
\textit{Transformer} & ::= & \lambda \texttt{a} . [E, E] \\
E  & ::= & \texttt{a.l}\ |\ \texttt{a.r}\ |\ 0\ |\ {-}E\ |\ \texttt{INTMAX}\ |\ \texttt{INTMIN}\  |\ \texttt{min}(E, E)\ |\ \texttt{max}(E, E)|\ \texttt{mul}(E, E)
\end{array}
\end{equation}

\eqref{unsignedAndTransformer} shows the abstract transformer synthesized by {\tool} for the unsigned-integer interval domain ($\uintv$).
In this case, the abstract transformer is quite simple, multiplying
the corresponding left and right limits of the multiplicands.

\begin{equation}
\label{Eq:unsignedAndTransformer}
\begin{array}{@{\hspace{0ex}}l@{\hspace{0ex}}}
\abst{\texttt{mul}}{no} \mathtt{(a_1: \uintv, a_2: \uintv) : \uintv}  = \mathtt{[mul(a_2.l,\ a_1.l),\ mul(a_2.r,\ a_1.r)]}
\end{array}
\end{equation}

\noindent
In contrast, the case of signed multiplication is more involved.
\eqref{signedAndTransformer} shows the abstract transformer synthesized by {\tool}
for the signed-integer interval domain ($\sintv$), again using the DSL from \eqref{LForIntervalDomainMul}.

\begin{equation}
\label{Eq:signedAndTransformer}
{
\begin{array}{@{\hspace{0ex}}l@{\hspace{0ex}}}
\abst{\mathtt{mul}}{no}\mathtt{(a_1: \sintv, a_2: \sintv) : \sintv = } \\
\quad \left[\begin{array}{@{\hspace{0ex}}l@{\hspace{0ex}}}
\mathtt{min\left(\begin{array}{@{\hspace{0ex}}l@{\hspace{0ex}}}
	\mathtt{min(mul(a_2.r,\ a_1.r),\ mul(a_1.r,\ a_2.l)),} \\
	\mathtt{min(mul(a_1.l,\ a_2.r),\ mul(a_2.l,\ a_1.l))}
	\end{array}\right)}, 
\mathtt{max\left(\begin{array}{@{\hspace{0ex}}l@{\hspace{0ex}}}
	\mathtt{max(mul(a_1.l,\ a_2.l),\ mul(a_2.r,\ a_1.r)),} \\
	\mathtt{max(mul(a_1.r,\ a_2.l),\ mul(a_1.l,\ a_2.r))}
	\end{array}\right)}
\end{array}\right]
\end{array}
}
\end{equation}

\forOOPSLA{
\begin{wrapfigure}{r}{.53\linewidth}
	\hspace{2mm}
}
\forARXIV{
\begin{figure}
}
	\begin{tabular}{c}
		\begin{lstlisting}[basicstyle=\small\ttfamily]
$\wintv$ xorGen ($\wintv$ a1, $\wintv$ a2) {
  $\wintv$ s1[] = intervalSplitAtZero(a1.l, a1.r);$\label{Li:splitAtZr1}$
  $\wintv$ s2[] = intervalSplitAtZero(a2.l, a2.r);$\label{Li:splitAtZr2}$
  $\wintv$ result = $\bot$;
  for(int i = 0; i < s1.size(); i++){ $\label{Li:loopOverSplits1}$
    for(int j = 0; j < s2.size(); j++){ $\label{Li:loopOverSplits2}$
      int res0 = $\fbox{?1}(s1[i],\ s2[j])$; $\label{Li:hole1}$
      int res1 = $\fbox{?2}(s1[i],\ s2[j])$; $\label{Li:hole2}$
      result = $Join_{\wintv}$([res0, res1], result); $\label{Li:joinw}$
}}}
		\end{lstlisting}
	\end{tabular}
	\caption{Sketch for $\mathtt{xor}^\sharp$ in $\wintv$ and $\sintv$. 
			Function \texttt{intervalSplitAtZero} splits the interval at 0, if interval contains 0.
		\label{Fi:xorSignedDSL}}
\forARXIV{
\end{figure}
}
\forOOPSLA{
	\vspace{-3mm}
\end{wrapfigure}
}
\noindent
This abstract transformer takes the product of every pair of bounds
from the two intervals, and chooses the minimum element as the left bound of the
resultant interval.
Similarly, it picks the maximum element as the right bound of the resultant interval.


Note that, for all the above domains, although {\tool} was provided
with the
same DSL syntax
shown in \eqref{LForIntervalDomainMul}, the supplied
semantics of the constructs differed, according to the domain.
That is, depending on the intended domain,
$\mathtt{a.l}$ and $\mathtt{a.r}$ were interpreted as signed or unsigned integers;
$\mathtt{INTMAX}$ and $\mathtt{INTMIN}$ were interpreted according to the bitwidth
and signedness under consideration;
etc.

\subsubsection{Abstract transformers for \texttt{xor}}
\label{Se:absTransformXor}

\twrchanged{
The abstract transformer for \texttt{xor} in the $\wintv$ domain
is quite complex because it involves nested loops.
However, the ``high-level'' algorithm is quite intuitive and proceeds by splitting intervals.
If $0^k$ falls in the interval $[a, b]$, where $a, b$ are $k$-bit bit-vectors,
a split at zero (South pole) in $\wintv$ will split the interval
$[a,b]$ into two intervals $[a, 1^k]$ and $[0^k, b]$.

We show the sketch for this high-level structure in \figref{xorSignedDSL},
where the the $i^{\textit{th}}$ \emph{hole} to be filled by {\tool} is denoted by ``\fbox{\texttt{?i}}''.
Both the multiplicands are split at 0
(\lineseqref{splitAtZr1}{splitAtZr2}), and the sketch loops through
each possible pair of interval segments generated
(\lineseqref{loopOverSplits1}{loopOverSplits2}).
For each pair, it specifies \textit{holes} for functions to be synthesized
(\lineseqref{hole1}{hole2}) that constitute the lower and upper bounds
of the interval.
Finally, the intervals corresponding to all such segments are joined
to construct the final interval (\lineref{joinw}).

{\tool} synthesizes the functions shown in \eqrefs{hole1}{hole2} for the two holes in \figref{xorSignedDSL},
using the following grammar:

\begin{align}
\label{Eq:DSLUXor}
\begin{array}{l}
  \textit{Transformer} ::= \lambda \mathtt{a_1, a_2} .\ EX \\
  \begin{array}{@{\hspace{0ex}}r@{\hspace{1.0ex}}c@{\hspace{1.0ex}}l@{\hspace{0ex}}}
    EX & ::=  & \texttt{minOr}(B,B,B,B) \mid \texttt{maxOr}(B,B,B,B) \mid \texttt{minAnd}(B,B,B,B) \mid \texttt{maxAnd}(B,B,B,B)\\
       & \mid & \texttt{minOr}(0, EX, 0, EX) \mid \texttt{maxOr}(0, EX, 0, EX) \mid \texttt{minAnd}(0, EX, 0, EX) \mid \texttt{maxAnd}(0, EX, 0, EX)\\
       & \mid & \texttt{or}(EX, EX) \mid \texttt{and}(EX, EX)\\
    B  & ::=  & a_1.l \mid a1_1.r \mid a_2.l \mid a_2.r \mid \mbox{\textasciitilde}B
  \end{array}
\end{array}
\end{align}




\begin{align}
\label{Eq:hole1}
{\mathtt{?1}\mathtt{(a_1: \wintv, a_2: \wintv) : \wintv}} = \mathtt{or}\left(\begin{array}{l}
\mathtt{minAnd}(a_1.l,a_1.r,\mbox{\textasciitilde}a_2.r,\mbox{\textasciitilde}a_2.l), \\
\mathtt{minAnd}(\mbox{\textasciitilde}a_1.r,\mbox{\textasciitilde}a_1.l,a_2.l,a_2.r)
\end{array}\right)
\end{align}

\begin{align}
\label{Eq:hole2}
{\mathtt{?2}\mathtt{(a_1: \wintv, a_2: \wintv) : \wintv}}  = \mathtt{maxOr}\left(\begin{array}{l}
0,\ \mathtt{maxAnd}(a_1.l,a_1.r,\mbox{\textasciitilde}a_2.r,\mbox{\textasciitilde}a_2.l), \\
0, \mathtt{maxAnd}(\mbox{\textasciitilde}a_1.r,\mbox{\textasciitilde}a_1.l,a_2.l,a_2.r)
\end{array}\right)
\end{align}

Using the sketch (\figref{xorSignedDSL}) and the same DSL (\eqref{DSLUXor}), with numbers and operations interpreted as signed integers {\tool} ends up synthesizing the same transformers for the signed domain ($\sintv$).
For the unsigned domain ($\uintv$), the transformer is simpler:
\tool could synthesize the $\uintv$ transformer just from the DSL
(\eqref{DSLUXor})---with numbers and operations interpreted as
unsigned---without any need for a sketch.
\eqref{uxormin} (\texttt{minXor}) and \eqref{uxormax}
(\texttt{maxXor}) show the minimum and maximum limits for the \texttt{xor}
abstract transformer in the $\uintv$ domain, respectively.

\begin{align}
\label{Eq:uxormin}
{\mathtt{minXor}\mathtt{(a_1: \uintv, a_2: \uintv) : \uintv}} = \mathtt{or}\left(\begin{array}{l}
\mathtt{minAnd}(a_1.l,a_1.r,\mbox{\textasciitilde}a_2.r,\mbox{\textasciitilde}a_2.l), \\
\mathtt{minAnd}(\mbox{\textasciitilde}a_1.r,\mbox{\textasciitilde}a_1.l,a_2.l,a_2.r)
\end{array}\right)
\end{align}

\begin{align}
\label{Eq:uxormax}
{\mathtt{maxXor}\mathtt{(a_1: \uintv, a_2: \uintv) : \uintv}} = \mathtt{maxOr}\left(\begin{array}{l}
0,\ \mathtt{maxAnd}(a_1.l,a_1.r,\mbox{\textasciitilde}a_2.r,\mbox{\textasciitilde}a_2.l), \\
0, \mathtt{maxAnd}(\mbox{\textasciitilde}a_1.r,\mbox{\textasciitilde}a_1.l,a_2.l,a_2.r)
\end{array}\right)
\end{align}

}

\smallskip
\noindent
\begin{mdframed}[innerleftmargin = 3pt, innerrightmargin = 3pt, skipbelow=-0.25em]
  \textbf{Finding [RQ1]:}
  The time taken by \tool to synthesize an $L$-transformer across all
  of the fixed-bitwidth interval-transformer experiments varies between 4.23s and 1,449.85s.
	\FuzzOut{
  \newline
  \textbf{Finding [RQ2]:}
		For each function, running tool with a logical specification and a closed-box specification of the function resulted in equivalent transformers.}
  \newline
  \textbf{Finding [RQ2]:}
  Our experiments using \tool uncovered bugs in the abstract
  transformers implemented by \citeauthor{APLAS:WrInterval12} for
  $\texttt{xor}$ for the $\sintv$ and $\wintv$ domains.
  The two bugs had a single root cause, which was
  that an auxiliary function used in their interval-analysis tool
  was unsound due to a mistranscription of code from
  \textit{Hacker's Delight}~\citep{hackersDelight},
  which came to light when we used the \citeauthor{APLAS:WrInterval12}
  auxiliary function, and \tool failed to synthesize a correct abstract transformer.

  After the bug in the auxiliary function was fixed,
  all of the synthesized abstract transformers for
  the three kinds of interval domains are sound, precise, and semantically
  equivalent to those provided in the implementation that accompanies the paper
  by \citeauthor{APLAS:WrInterval12}.
\end{mdframed}

\subsection{Experience with Designing DSLs}
\label{Se:ExperienceWithDesigningDSLs}


\twrchanged{
In this section, we discuss more about the design of DSLs for transformer synthesis,
using the DSLs shown in \eqrefsp{DSLForContainsCI}{DSLForTrimCI}{TrimPSDSL} as examples.

Some aspects of DSLs are common across the DSLs for different operations and different
abstract domains.
For instance, all the DSLs in \eqrefsp{DSLForContainsCI}{DSLForTrimCI}{TrimPSDSL}
provide a way to check whether an abstract value is $\bot$ or $\top$, and
to perform different actions depending on the outcome.

Other parts of a DSL typically reflect the properties that are
observable in the abstract domain for which the DSL will be used for
synthesizing transformers.
For instance, in the case of the DSL for the \texttt{contains} operator in
the character inclusion ($\charin$) domain (\eqref{DSLForContainsCI}),
it is natural to add constructs such as \texttt{isSubset()} and \texttt{isEmpty()}
to compare sets of characters.
In the case of the DSL for the \texttt{trim} operation in $\charin$
domain (\eqref{DSLForTrimCI}), we use constructs such as \texttt{containsSpace()}
and \texttt{removeSpace()} to handle the space (\texttt{\textvisiblespace}) character.
Similarly, \eqref{TrimPSDSL} is also a DSL for the \texttt{trim} operation, but for
use with the prefix-suffix ($\mathcal{PS}$) domain.
\eqref{TrimPSDSL} is similar to that of \eqref{DSLForTrimCI}, but instead of operations on sets,
here operations are on strings, such as \texttt{trimStart()} and \texttt{trimEnd()}, which
return strings in which spaces have been removed from the beginning and the end, respectively,
of the argument string.

We now turn to what we observed when \tool is supplied with a DSL that
is a misfit for the problem at hand.
Such misfitting can take two forms:
one can have ``too many'' constructs (\sectref{TooManyConstructs}), or
``too few'' constructs (\sectref{TooFewConstructs}).

\subsubsection{DSLs with ``too many'' constructs}
\label{Se:TooManyConstructs}

Consider the DSL shown in \eqref{DSLTooMuch}.

\begin{equation}
  \label{Eq:DSLTooMuch}
  \begin{array}{@{\hspace{0ex}}r@{\hspace{1.0ex}}c@{\hspace{1.0ex}}l@{\hspace{0ex}}}
    \textit{Transformer} & ::= & \lambda \texttt{a} . [E, E] \\
    E & ::= & \texttt{a.l} \mid \texttt{a.r} \mid  0 \mid {-}E \mid {+}\infty \mid {-}\infty \mid E + E \mid E - E \mid E * E \\
      &     & \mid \texttt{min}(E, E) \mid \texttt{max}(E, E) \mid \hl{\texttt{E * E * E}}  \mid \hl{\texttt{pow(E, E)}}
  \end{array}
\end{equation}

\forOOPSLA{
\begin{wrapfigure}{r}{.54\linewidth}
}
\forARXIV{
\begin{figure}
}
	\hspace{1mm}
	\begin{tabular}{c}
\begin{lstlisting}[language=C,numbers=left]
$\texttt{contains}^\sharp_\texttt{syn}(\mathtt{a_1}:CI) (\mathtt{a_2}:CI): \mathtt{AbsBool} =$
 ite(isBot($\mathtt{a_1.l,a_1.u}$) $\lor$ isBot($\mathtt{a_2.l,a_2.u}$),
   boolBot,
   ite(isTop($\mathtt{a_1.l,a_1.u}$) $\land$ isSubset($\mathtt{a_2.u, a_1.l}$),$\label{Li:woEmpTrueCase}$
     boolTrue,$\label{Li:retTrueForWoEmpCi}$
     ite($\neg$isSubset($\mathtt{a_2.l,a_1.u}$),
       boolFalse,
       boolTop )))
\end{lstlisting}
	\end{tabular}
	\caption{Another abstract transformer for \texttt{contains} in the $\mathcal{CI}$ domain.\label{Fi:CiConNotPrec}}
\forARXIV{
	\end{figure}
}
\forOOPSLA{
	\vspace{-2mm}
\end{wrapfigure}
}

\noindent
\eqref{DSLTooMuch} is similar to the DSL shown in \eqref{LForIntervalDomain},except that it
has some extra constructs, which are highlighted in yellow.
We tried to synthesize an abstract transformer for \texttt{abs} with this DSL variant.
We ran \tool three times with a timeout of 6,000 seconds
per call on Sketch---ten times the usual timeout.
The median time of the three runs was 6010.7 seconds.
In all three runs, \tool could only synthesize a sound but imprecise
interval transformer for absolute value in which the right-hand limit
of the return value is always positive infinity ($+\infty$).
An investigation of the reason for this result revealed that \tool
exceeded the timeout threshold in calls to \func{CheckPrecision},
causing it to synthesize an imprecise solution.

\subsubsection{DSLs with ``not enough'' constructs}
\label{Se:TooFewConstructs}

Now consider the DSL defined by \eqref{DSLForContainsCI} but with the
production $B~::=~\texttt{isEmpty}(CI)$ removed, and suppose that we
ask \tool to synthesize an abstract transformer for the
\texttt{contains} operation in the $\mathcal{CI}$ domain.
In this case, \tool synthesizes the abstract transformer shown in
\figref{CiConNotPrec}.
This abstract transformer is a best $L$-transformer, where $L$ is the
language defined by \eqref{DSLForContainsCI} without the production
$B~::=~\texttt{isEmpty}(CI)$.

The absence of \texttt{isEmpty()} in the DSL causes
\figref{CiConNotPrec} to be less precise than the (corrected) abstract
transformer shown in \figref{containsSafe}
(without \lineref{BuggyCondition} and with \lineref{FixIsEmpty}).
Concretely, the empty string is contained in every string.
The corrected abstract transformer in \figref{containsSafe}
returns \texttt{boolTrue} whenever $a_2$ is the empty
string---see \figref{containsSafe}, \lineref{retTrue}.
In contrast, the transformer in \figref{CiConNotPrec} returns \texttt{boolTrue}
only in the case that $\mathtt{a_2}$ is the empty string and 
$\mathtt{a_1}$ is $\top$:
in \lineref{woEmpTrueCase} it checks whether $a_1$ is $\top$ and $a_2.u$ is subset of $a_1.l$;
when $a_1$ is $\top$ and $a_2$ is the empty string, $a_1.l$ and $a_2.u$ both hold the empty set,
and \texttt{boolTrue} is returned (\lineref{retTrueForWoEmpCi}).
However, when $a_1$ is not $\top$, due to the absence of \texttt{isEmpty()} in the DSL,
the transformer cannot check whether argument $a_2$ is the empty string, and thus returns
the sound answer \texttt{boolTop}.
}


\section{Related Work}
\label{Se:RelatedWork}

The related work closest to ours was discussed  in \sectref{Introduction};
here we discuss some other related work.

Program synthesis has recently gained a lot of attention,
and has found applications in diverse areas.
CEGIS~\citep{Lezama13} is a popular synthesis strategy.
In our work, we interleave two CEGIS loops to handle competing objectives,
considering a ``negative-example'' classification as a soft constraint,
allowing \texttt{MaxSatSynthesize} to find a better classification.
Work on synthesizing data-structure invariants
\citep{PLDI:Anders20} also deals with two
competing objectives---weakening/strengthening candidate invariants---for which
they employ three CEGIS loops.


\twrchanged{
Several papers by Reps, Sagiv, Yorsh, Thakur, and others address
(explicitly or implicitly) the problem of creating best abstract
transformers for a variety of abstract-interpretation frameworks
\cite{VMCAI:RSY04,DBLP:conf/cav/ThakurR12,DBLP:conf/sas/ThakurER12,ThakurLLR15,VMCAI:Reps16}
(with slightly different requirements among the different papers).
\citet{ELSAR14} gave a method for creating best abstract transformers
for the abstract domain of conjunctions of bit-vector equalities.
The main differences with our work is that (i) those papers use
positive examples only, and (ii) they do not allow the user to specify a DSL.
Our algorithm uses both positive and negative examples, and the user
can supply a DSL of their own design.
}

\citet{CAV:Wang18} presented an approach for learning
abstract transformers for a given abstract domain.
There are major differences between their approach and ours.
For them, (i) abstract transformers are expressed in a specific language:
conjunctions of a (learned) set of fixed predicates over affine expressions,
and (ii) the operations of the DSL are the concrete operations for which
the system tries to find suitable abstract transformers.
In contrast, with \tool the user (i) supplies their own DSL in which
the abstract transformer is to be expressed, and
(ii) provides a logical specification of the concrete operation
for which an abstract transformer is sought.

\citet{bielik2017} present a method to learn program analyzers from data.
Their method attempts to automatically learn program-analysis inference
rules for program primitives (such as assignment statements, pointer
dereferences, etc.), with an emphasis on learning corner-cases for
such rules.
The user must supply training data of the form
$\langle \textit{program}, \textit{analysis output} \rangle$.
The algorithm finds patterns to apply to program abstract-syntax trees
to produce analysis results that match the dataset as closely as possible.
It also uses program-mutation operations to test the learned rules,
and to augment the dataset in a CEGIS loop.
Transfer functions are expressed as decision trees, learned using a
modification of the ID3 algorithm.
Because the transfer functions cannot use
arithmetic/bitwise operations, their method is suited to
\textit{selecting} from a set of facts, rather than
\textit{constructing} arbitrary abstract values.
\citeauthor{bielik2017} handle precision by attempting to minimize a
cost function on the dataset.
They do not attempt to verify that their solution is indeed precise.
While their approach works well for pointer analysis, and for some forms of
type analysis and constant propagation, due to (i) the inability to
use arithmetic, and (ii) the way precision is handled, their technique
cannot generate abstract transformers for the domains
considered in our experiments.

A recent paper~\citep{ICSE:Wang21} is a synthesis-based technique for
creating sound abstract transformers using learned predicates;
however, their method of using Datalog query containment corresponds
to \func{CheckSoundness} only.
They have no analogue of \func{CheckPrecision}, and hence no mechanism for
controlling the precision of the transformers that they obtain.

\citet{PLDI:PF21} synthesize code specifications using CHC solvers. The problem they tackle shares some commonalities with ours: a specification has to be not only sound but precise (e.g., True is not a very useful specification). To synthesize precise specifications, their algorithm uses CHC solvers to strengthen the synthesized specification in a CEGIS loop. While this aspect shares some structure with our precision queries, the task solved by our algorithm is much harder as it requires synthesizing programs over a DSL instead of logical specifications in a given theory.  Moreover, the problem that they address is dual to ours: their work goes from code to logic, whereas our work goes from logic to code.

\twrchanged{
\citet{OOPSLA:Astorga21} use a test generator to
generate positive examples to synthesize a contract that is sound with
respect to the examples.
There are two main differences between their work and ours.
(i) They only use positive examples, whereas we use both positive and
negative ones;
negative examples are the key to synthesizing best $L$-transformers.
(ii) Their notion of ``tight'' is with respect to a syntactic restriction
on the logic in which the contract is to be specified.
That is, their system works with a specific logic fragment---for
example, the contract is to be specified by a formula that uses at
most $k$ disjuncts.
In contrast, in our work the user-specified DSL provides another
``knob,'' which can be used to explore the trade-off between precise
solutions and pragmatic solutions. There has also been the use of test generators (like fuzzers) to analyze programs for \textit{closed-box} functions (program components whose logical specifications are not available)~\cite{ISSTA:PandeyKR19,ISSTA:LR22}. \tool can also be targeted for such applications to infer abstract transformers for such \textit{closed-box} functions by driving the soundness check by a test generator; we intend to pursue such directions in the future.
}


\section{Conclusion}
\label{Sec:conclusion}
This paper presents an algorithm for automatically creating abstract
transformers by means of program synthesis.
Given a concrete operation $\textit{op}$, an abstract domain $\AbsDomain$, and
a DSL $L$ in which the abstract transformer for $\textit{op}$ is to be expressed,
our algorithm is guaranteed to produce a best $L$-transformer for $\textit{op}$.
The synthesis algorithm itself is novel, in that it uses Occam's razor
to overcome the situation of there being a conflict between examples that
are (currently) labeled ``positive'' and ``negative.''
In this case, the goal of the synthesis step becomes ``synthesize an
answer that ignores the smallest number of negative examples.''
The algorithm should have applications in other kinds of synthesis problems
as a way of accommodating inductive bias.

We incorporated the algorithm into a tool, called \tool, and used it to study
the performance of our approach on fifteen concrete operators and
eight abstract domains:
five string domains and three fixed-bitwidth interval domains.
During our experiments, we discovered four soundness bugs in the manually written
transformers used in abstract-interpretation engines that employed these domains,
showcasing the value of our approach.
Furthermore, our synthesized transformers match the manually written
transformers in terms of both precision and performance.


\begin{acks}
Supported, in part,
by a gift from Rajiv and Ritu Batra;
by multiple Facebook Research Awards;
by a Microsoft Faculty Fellowship;
by NSF under grants 1420866, 1763871, 1750965, 1918211, and 2023222; 
and by ONR under grants N00014-17-1-2889 and N00014-19-1-2318.
Any opinions, findings, and conclusions or recommendations
expressed in this publication are those of the authors,
and do not necessarily reflect the views of the sponsoring
entities.
\end{acks}

\bibliographystyle{ACM-Reference-Format}
\bibliography{refs}



\end{document}


\maketitle
    \section{$\mathtt{concat}^\sharp$ transformer for $\mathcal{SH}$ domain}
    The sketch of the target abstract transformer $\mathtt{concat}^\sharp$ in $\mathcal{SH}$ domain from \textbf{Case Study-1} ($\S.6.1$ of the main paper) is shown in \figref{concatSk}. The target transformer is complex as it requires a loop. 
    
    In the sketch, for a set of choices over expressions, $\{\mid e_1 \mid e_2 \mid \dots e_n \mid\}$, {\tool} selects the ``right" expression $e_i$ for all the corresponding \textit{hole}. Overall, it selects the right candidates from the choices corresponding to each hole in an attempt to synthesize a best L-transformer. For example in Line~\ref{Li:HashDSLChoice}, one of the expressions from \texttt{rotateLeft(r,1)} or \texttt{rotateRigth(r,1)} is selected for assignment to variable \texttt{r}. The procedure \texttt{reverse()} reverses the order of bits in input bitvector, \texttt{rotateLeft(bv, k)} is circular shift left, and \texttt{rotateRight(bv, k)} performs circular shift right of input bitvector \texttt{bv} by  \texttt{k} bits.

    \begin{figure}[H]
\begin{lstlisting}[numbers=left]
#define W 64

bit[W] absConcat(bit[W] a, bit[W] b){
  bit[W] r = {| a | b | reverse(a) | reverse(b)|};
  bit[W] one = {1};
  bit[W] c = {0};
  bit[W] cond;
  for(int i = 0; i < W; i++){
    one = {1};
    r = {| rotateLeft(r, 1) | rotateRight(r, 1) |}; $\label{Li:HashDSLChoice}$
    cond = {| (a & r) | (a | r) | (a ^ r) |};
    if( (cond) != (bit[W]){0}){
      c = {| (c & rotateLeft(one, i))
           | (c | rotateLeft(one, i))
	   | (c ^ rotateLeft(one, i))
	   | (c & rotateRight(one, i))
	   | (c & rotateRight(one, i))
	   | (c ^ rotateRight(one, i)) |};
    }
  }
  return c;
}
\end{lstlisting}
        \caption{Sketch to synthesize abstract transformer for $\mathtt{concat}^\sharp$ in $\mathcal{SH}$\label{Fi:concatSk}}
    \end{figure}

    \vspace{20pt}
    
    With the above sketch, {\tool} is able to synthesize a sound and precise abstract transformer shown in \figref{concatAmurth}.

    \begin{figure}
        \begin{lstlisting}[numbers=left]
#define W 64

bit[W] absConcat(bit[W] a, bit[W] b) {
    bit[W] r = reverse(b);
    bit[W] one = {1};
    bit[W] c = {0};
    bit[W] cond;

    for(int i = 0; i < W; i++){
        one = {1};
        r = rotateLeft(r, 1);
        cond = a & r;
        if( (cond) != (bit[W]){0}){
            c = (c ^ rotateLeft(one, i));
        }
    }
    return c;
}
        \end{lstlisting}
        \caption{Abstract transformer synthesized by {\tool} for $\mathtt{concat}^\sharp$\label{Fi:concatAmurth}}
    \end{figure}